\documentclass{emulateapj}
\usepackage{times}
\usepackage{hyperref}

\newcommand  \nii   {[N{\sc ii}\relax]}

\newcommand{\beq}{\begin{equation}}
\newcommand{\eeq}{\end{equation}}
\def\alp{\mbox{$\alpha$}}

\def\earth{\hbox{$\oplus$}}
\def\arcmin{\hbox{$^\prime$}}
\def\arcsec{\hbox{$^{\prime\prime}$}}

\newcommand{\AmS}{{\protect\the\textfont2
  A\kern-.1667em\lower.5ex\hbox{M}\kern-.125emS}}
\newcommand{\lsim}{\ \raise
-2.truept\hbox{\rlap{\hbox{$\sim$}}\raise5.truept\hbox{$<$}\ }}
\newcommand{\gsim}{\ \raise
-2.truept\hbox{\rlap{\hbox{$\sim$}}\raise5.truept\hbox{$>$}\ }}
\newcommand{\simsim}{\ \raise
-2.truept\hbox{\rlap{\hbox{$\sim$}}\raise5.truept\hbox{$\sim$}\ }}

\def\Q{\ifmmode\mathcal{Q}\else$\mathcal{Q}$\fi}

\slugcomment{Accepted by the Astrophysical Journal January 13, 2012}

\righthead{Clustering Behavior of PMS stars in NGC 602}
\lefthead{Gouliermis, et al.}

\shorttitle{Clustering Behavior of PMS stars in NGC 602}
\shortauthors{Gouliermis, et al.}

\begin{document}

\title{The clustered nature of star formation. Pre--main-sequence clusters in the star-forming 
region NGC~602/N90 in the Small Magellanic Cloud\altaffilmark{1}}

%\title{Clustered Star Formation in the region NGC~602/N90 in the Small 
%Magellanic Cloud\altaffilmark{1}}

%\title{Clustered Star Formation in the region NGC~602/N90 in the Small 
%Magellanic Cloud observed with the Hubble Space Telescope\altaffilmark{1,}\altaffilmark{2}}

%\title{Clustering of Pre-Main Sequence Stars in the 
%SMC star-forming region NGC~602/N90\altaffilmark{1,}\altaffilmark{2}}

%\footnotemark[1]
%\footnotetext[1]{}


\author{Dimitrios A. Gouliermis\altaffilmark{2, 3}, 
        Stefan Schmeja\altaffilmark{4, 3}, 
        Andrew E. Dolphin\altaffilmark{5}, 
        Mario Gennaro\altaffilmark{2},\\
        Emanuele Tognelli\altaffilmark{6,7},
        and 
        Pier Giorgio Prada Moroni\altaffilmark{6,7} 
        }

\altaffiltext{1}{Based on observations made with the NASA/ESA {\em 
Hubble Space Telescope}, obtained at the Space Telescope Science 
Institute, which is operated by the Association of Universities for 
Research in Astronomy, Inc. under NASA contract NAS 5-26555.}

%\altaffiltext{2}{Research supported by the German Aerospace 
%Center (DLR) and the German Research Foundation (DFG).}

\altaffiltext{2}{Max Planck Institute for Astronomy, K\"{o}nigstuhl 
17, 69117 Heidelberg, Germany}

\altaffiltext{3}{Zentrum f\"ur Astronomie der Universit\"at Heidelberg, 
Institut f\"ur Theoretische Astrophysik, Albert-Ueberle-Str.~2, 69120 
Heidelberg, Germany}

\altaffiltext{4}{Zentrum f\"{u}r Astronomie der Universit\"{a}t Heidelberg,
Astronomisches Rechen-Institut, M\"{o}nchhofstr. 12-14, 69120 
Heidelberg, Germany}

\altaffiltext{5}{Raytheon Company, PO Box 11337, Tucson, AZ 85734, USA}

\altaffiltext{6}{Dipartimento di Fisica ``Enrico Fermi'',  Universit\`{a} 
di Pisa, largo Pontecorvo 3, Pisa I-56127, Italy} 
 
\altaffiltext{7}{INFN-Sezione di Pisa, largo Pontecorvo 3, Pisa I-56127, Italy}

%Department of Physics "E.Fermi"
%Largo Bruno Pontecorvo 3, 56127, Pisa, Italy

%\email{dgoulier@mpia-hd.mpg.de}

%\email{sschmeja@ari.uni-heidelberg.de}

\begin{abstract}

Located at the tip of the wing of the Small Magellanic Cloud (SMC), the star-forming region NGC~602/N90 
is characterized by the HII nebular ring N90 {and the young cluster of pre--main-sequence (PMS) and 
early-type main sequence stars} NGC~602, located in the central area of the ring. We present a thorough cluster analysis of the  
stellar sample identified with HST/ACS camera in the region.  We show that apart from the 
central cluster, low-mass PMS stars are congregated {in thirteen additional small compact sub-clusters} at the periphery of 
NGC~602, identified in terms of their higher stellar density in respect to the average background density 
derived from star-counts. We find that the spatial distribution of the PMS stars is bimodal, {with 
an unusually large fraction ($\sim$~60\%) 
of the total population being clustered}, while the remaining is diffusely distributed in the inter-cluster area, 
covering the whole central part of the region. From the corresponding color-magnitude diagrams (CMDs) 
we disentangle an age-difference of $\sim$~2.5~Myr between NGC~602 and the compact sub-clusters 
which appear younger, on the basis of comparison of the brighter PMS stars with evolutionary 
models, which we accurately calculated for the metal abundance of the SMC. The diffuse PMS population 
appears to host stars as old as those in NGC~602. Almost all detected PMS sub-clusters appear 
to be centrally concentrated. When the complete PMS stellar sample, including both clustered and diffused 
stars, is considered in our cluster analysis  it appears as a single centrally concentrated stellar agglomeration, 
covering the whole central area of the region. Considering  also the hot massive stars of the system, we find evidence that this agglomeration 
is hierarchically structured.  Based on our findings we propose a scenario, according to which the region NGC~602/N90 
experiences an active clustered star formation for the last $\sim$~5~Myr. The central cluster NGC~602 was 
formed first and {rapidly started dissolving into its immediate ambient environment}, possibly ejecting also massive stars found away from its
center. Star formation continued in sub-clusters of a larger stellar agglomeration, introducing 
an age-spread of the order of 2.5~Myr among the PMS populations. 
%This clustering behavior resembles that of a typical young stellar association {in the making.} 

%If we also consider previous indications from mid-IR observations that 
%star formation in some of the sub-clusters is on-going, then this stellar system resembles more a stellar association 
%on the making.

%Star formation continues taking place in compact PMS 
%sub-clusters at the periphery of NGC~602 with a diffuse inter-cluster PMS 
%population being originated from the clusters dissolution, introducing an age-spread among the various PMS populations of the order of 2.5~Myr.

\end{abstract}


\section{Introduction}

Stars are  usually born in groups \citep[e.g.][]{lada03, sta+pal05} in a large variety of 
spatial scales, from small compact clusters to large loose stellar complexes 
\citep{efelm98}, related to each other in a hierarchical fashion.  This hierarchy 
in the formation of stellar groups is thought to be {inherited from the complex hierarchical}
structure of the interstellar medium \citep[ISM; e.g.][]{maclow+klessen04, 
mckee+ostriker07}. The most likely source of this hierarchy is a combination of 
agglomeration  with fragmentation \citep{mclaughlin96}, self-gravity \citep{goodman09}, 
and turbulence \citep[][]{elmegrscalo04}. 
Stellar structures have hierarchical patterns not only in space, but also in time, 
suggesting that their formation takes place faster in the densest regions of a turbulent 
ISM \citep{elmegreen10}. 

%Such patterns of sub-clustering are observed in embedded 
%star-forming regions of our Galaxy \citep[e.g.,][]{skf08, gutermuth09}.

%The investigation of the clustering properties of extragalactic star-forming 
%regions can be facilitated only by the high resolving efficiency of {\sl Hubble 
%Space Telescope} (HST) and therefore is constrained to less embedded 
%regions. 

The Magellanic Clouds (MCs), the nearest galactic companions 
to the Milky Way, are the best templates of extragalactic star formation
due to their closeness to us and their location away from the
obscuring Galactic disk. The investigation of star-forming regions in the 
MCs has been significantly benefitted by the high resolving efficiency 
over a wide field-of-view provided by the {\sl Hubble Space Telescope} (HST). 
In particular, recent observations with the {\sl Wide-Field Planetary Camera 2} (WFPC2) and 
the {\sl Advanced Camera for Surveys} (ACS) on board HST provide a unique 
opportunity to investigate extragalactic clustered star formation with high spatial
resolution and sensitivity in wide-field coverage that corresponds to
physical length-scales between $\sim$~10 and 100~pc.

%\clearpage 
%%%%%%%%%%%%%%%%%%%%%%%%%%%% FIGURE %%%%%%%%%%%%%
\begin{figure*}[t!]
\centerline{\includegraphics[angle=0,clip=true,width=0.775\textwidth]{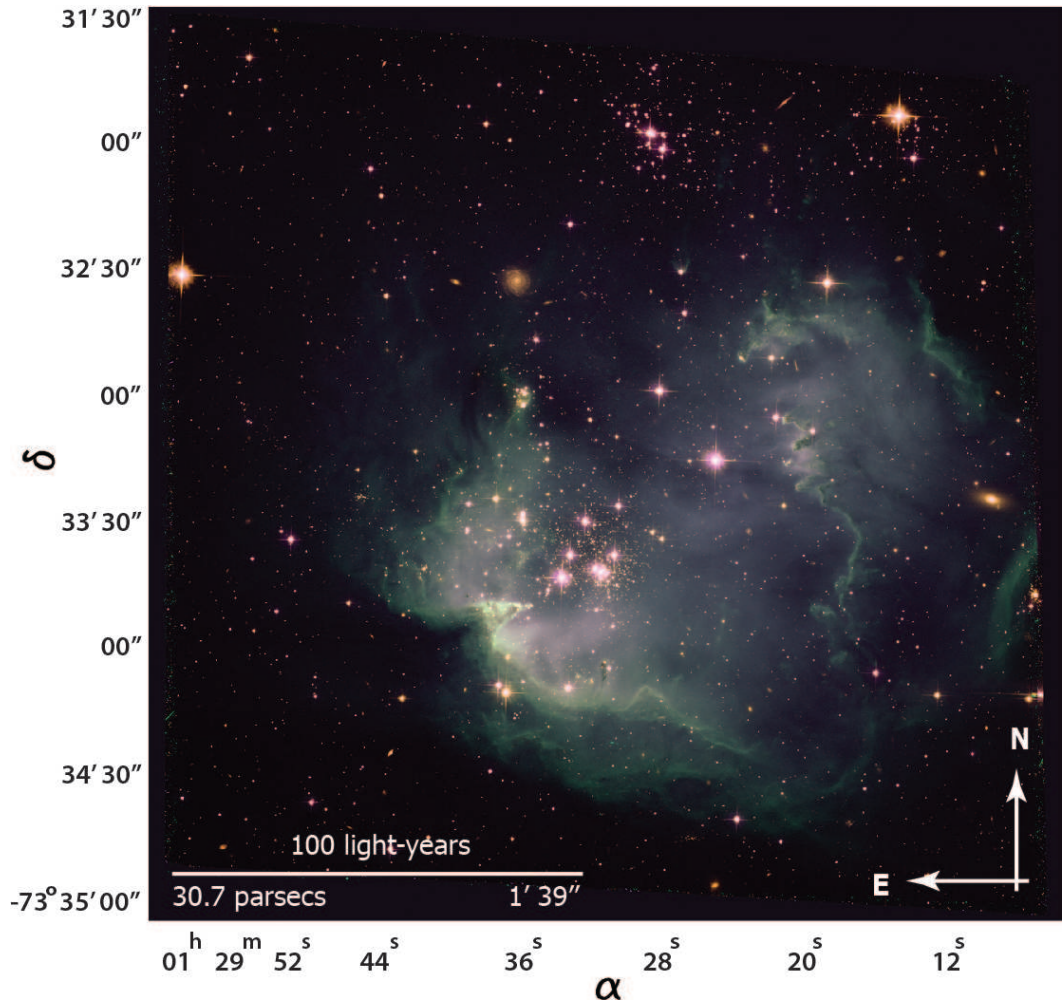}}
\caption{Color composite image of exposures made by ACS/WFC in 
two broadband filters, F555W ($V$ ) and F814W ($I$ ), and one 
narrowband filter, F658N (H{\alp} \&  {\nii}) of the region NGC~602/N90. 
The young cluster NGC~602 lies inside the ring-shaped nebula N90, 
the edges of which are eroded by the radiation of the hot young stars, producing 
the spectacular bubble. Ridges of dust and gaseous filaments are seen towards the 
northwest and the southeast. Dusty pillars point towards the center giving signs of the 
eroding effect by the central hot blue stars. Candidate YSOs found along 
the surrounding dust ridges give evidence that star formation has possibly started at 
the center of the region and advanced outward. In this study PMS stars are found to be 
distributed in a diffuse manner centered on NGC~602, as well as concentrated 
in younger compact peripheral sub-clusters along the inner rim of the bubble, providing 
support to the above suggestion.  \label{f:pic}}
\end{figure*}
%%%%%%%%%%%%%%%%%%%%%%%%%%%%%%%%%%%%%%%%%%%%%%

In our recent study of the stellar clustering behavior  of the 
brightest star-forming region in the Small Magellanic Cloud 
(SMC), NGC~346/N66, we found that it 
is hierarchically structured and fragmented into ten individual
subgroups of pre-main sequence (PMS) stars \citep[][]{schmeja09}.
The interplay between gravoturbulent cloud fragmentation 
and early dynamical evolution seems to be the main mechanism for 
building up large star-forming clusters 
\citep[e.g.][]{klessen00, clark05b}, and our study suggests that this
mechanism can best explain our findings in the region NGC~346/N66.
Continuing our work on the understanding of clustered
star formation in the SMC, we investigate here another star-forming region, 
NGC~602/N90 (Fig.~\ref{f:pic}). 

This region, located in the wing of the SMC, seems to
represent a different, less complex, mode of star formation than that of 
NGC~346/N66 \citep[see also][]{gouliermis07, gouliermis08}.  The emission
nebula LHA~115-N90 \citep{henize56}, or in short N90\footnote{N90 is also 
known as DEM~S~166 \citep{davies76}}, is a ring-shaped {\sc Hii} region with the
bright young cluster NGC~602 {\citep{westerlund64, hodge85}} located in its cavity. 
Being quite isolated and remote from the main body of the SMC this rather small star-forming 
region {(with a physical size of $\sim$~60~pc)} has
attracted interest in various investigations due to its brightness. Its
stellar population is known to {comprise massive hot stars younger than
15~Myr \citep[e.g.][]{battinelli92, massey00}}, with the eleven brightest
stars being early-type dwarfs with spectral types between O6 and B5
\citep{hutchings91}.

{The metallicity and dust-to-gas ratio} of NGC~602/N90 does not seem to differ
from typical values for the SMC, being [Fe/H]~$\simeq-$0.65 
\citep[e.g.][]{rolleston99, lee05}, and $\sim$~1/30 of that in the 
Milky Way \citep{stanimirovic00} respectively. The region is presumed 
to have been formed in a relatively isolated and diffuse environment 
by a star formation event that was produced by compression and turbulence 
associated with {\sc Hi} shell interactions, started at $\sim$~7~Myr ago,  
leading to the formation of stars $\sim$~3~Myr later, i.e., 4~Myr ago \citep{nigra08}.
However,  observations with the {\sl Spitzer Space Telescope} (STT) 
show that the region still hosts ongoing star formation demonstrated by a 
number of candidate massive Young Stellar Objects \citep[YSOs;][]{carlson07,
gouliermis07, carlson11}. Moreover, imaging with HST/ACS of the region 
revealed a plethora of candidate PMS stars with masses reaching the 
sub-solar regime \citep{carlson07, schmalzl08, cignoni09}. 

Preliminary studies of the spatial distribution of these stars show that they are not
homogeneously distributed across the region, but they are grouped 
in discrete compact concentrations. Some of those compact sub-clusters 
at the periphery of the bubble of N90 are found to coincide with candidate 
YSOs \citep{gouliermis07, carlson11}, suggesting that they are
probably still embedded.
In this paper we perform a thorough cluster analysis on 
the rich PMS population unveiled by HST/ACS in the bright SMC star-forming
region NGC~602/N90 (Fig.~\ref{f:pic}). The aim is to analyze the stellar 
clustering behavior of the region and to compare our
results to current scenarios of star formation in order to
achieve a more comprehensive understanding of clustered star formation
in the isolated environment of the SMC wing. 

We make use of our deep ACS photometry \citep{schmalzl08}, {and we 
apply the star-counts algorithm} to isolate local stellar density enhancements 
from the underlying density background and to identify all 
PMS clusters in the region. The dataset is presented in \S~\ref{sec:data}. The stellar
content of the region and  the thorough selection of the PMS population
are described in \S~\ref{s:stars}. The detection of individual young
stellar clusters in the region with the application of {\sl star-counts}, and the
extraction of their structural parameters is performed in \S~\ref{s:clusdet},
where also the stellar content of the individual sub-clusters and their surrounding
area is investigated. The clustering behavior of  the young
stellar content of the region is further quantified with the application of
the {\sl minimum spanning tree} method, which takes place in  \S~\ref{s:mst}.
{In \S~\ref{s:scenario} we discuss our results in terms of the star clusters formation
process as it is revealed from our analysis, and propose a scenario that
may sufficiently explain this process in the region. We give
our concluding remarks in \S~\ref{sec:discussion}.}

%\clearpage 
%%%%%%%%%%%%%%%%%%%%%%%%%%%% FIGURE %%%%%%%%%%%%%
\begin{figure}[t!]
\centerline{\includegraphics[angle=0,clip=true,width=0.875\columnwidth]{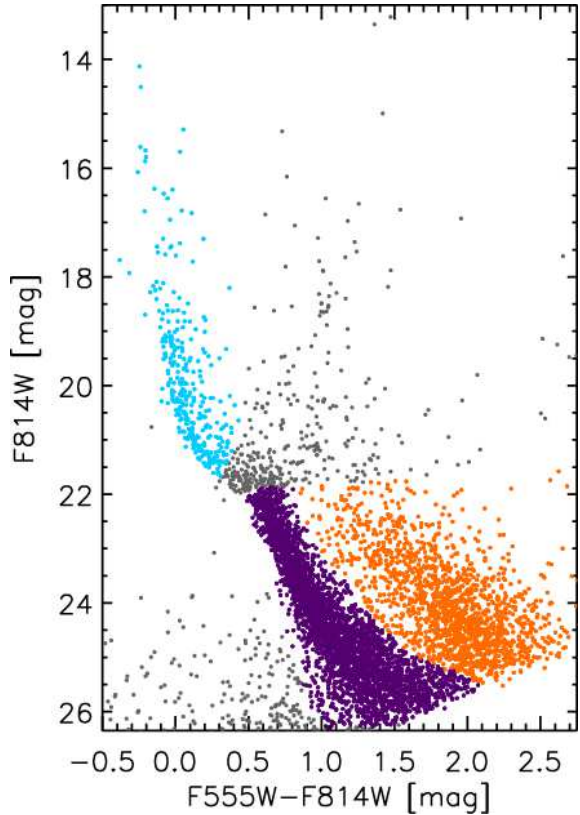}}
\caption{F555W$-$F814W, F814W CMD of all stars found in the 
region NGC~602/N90 with ACS/WFC imaging. The members of the three 
dominant stellar types, i.e, UMS, LMS and PMS stars (see \S~\ref{s:stars}), 
are plotted with light-blue, purple and orange colors respectively.  \label{f:cmd}}
\end{figure}
%%%%%%%%%%%%%%%%%%%%%%%%%%%%%%%%%%%%%%%%%%%%%%

\section{Observations and Photometry}
\label{sec:data}

The observations used in this study are taken with the Wide-Field Channel
(WFC) of ACS within the HST GO Program 10248 (PI: A. Nota). The HST 
images of one pointing, centered on the young cluster 
NGC~602, observed in the filters F555W ($\equiv V$) and F814W 
($\equiv I$),  were retrieved from the HST Data Archive\footnote{The HST Data 
Archive is accessible from MAST at\\ {\tt \url{http://archive.stsci.edu/hst/}}, and 
ESO at\\ {\tt \url{http://archive.eso.org/cms/hubble-space-telescope-data}}.}. These images cover an area of 
about $3.4\arcmin \times 3.5\arcmin$ ($\sim 58~{\rm pc} \times 58~{\rm pc}$ at the 
distance of the SMC). The reduction and photometry of the data is 
{discussed in detail in \cite{schmalzl08},} within our investigation of the Initial Mass 
Function in the region.  Photometry was performed using the ACS module of the package 
{\sc dolphot}, an adaptation of the photometry package HSTphot 
\citep{dolphin00}, and after eliminating bad detections more than 
5\,626 stars down to $m_{\rm 555} \simeq 26.5$~mag were included 
in the photometric catalog, with a completeness of \gsim~50\% down 
to $m_{\rm 555} \approx 25$~mag. The corresponding $VI$-equivalent 
color-magnitude diagram (CMD) is shown in Fig.~\ref{f:cmd}. 

%\clearpage 
%%%%%%%%%%%%%%%%%%%%%%%%%%%% FIGURE %%%%%%%%%%%%%
\begin{figure}[t!]
\centerline{\includegraphics[angle=0,clip=true,width=\columnwidth]{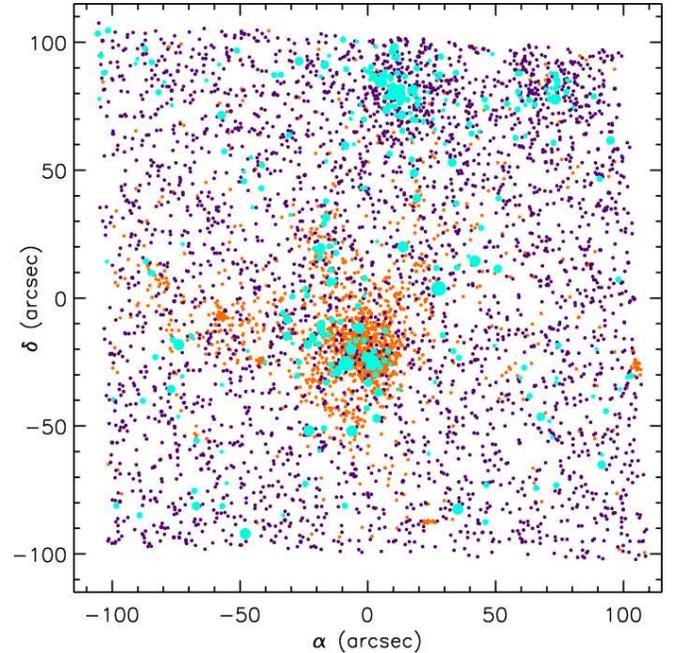}}
\caption{The stellar chart of stars identified with our photometry in the 
observed WFC pointing. Stars are color-coded according to their types, as derived
from their CMD positions; light-blue for UMS, purple for LMS, and 
orange for PMS stars (see text in \S~\ref{s:pmss}). 
{Larger blue symbols correspond to brighter UMS stars. Turn-off stars are not plotted.} 
The map is centered according to the observed 
WFC pointing (J2000 22:21:53.55 $-$73:33:17.14).  \label{f:map}}
\end{figure}
%%%%%%%%%%%%%%%%%%%%%%%%%%%%%%%%%%%%%%%%%%%%%%

%\clearpage 
%%%%%%%%%%%%%%%%%%%%%%%%%%%% FIGURE %%%%%%%%%%%%%
\begin{figure*}[t!]
\centerline{\includegraphics[angle=0,clip=true,width=0.775\textwidth]{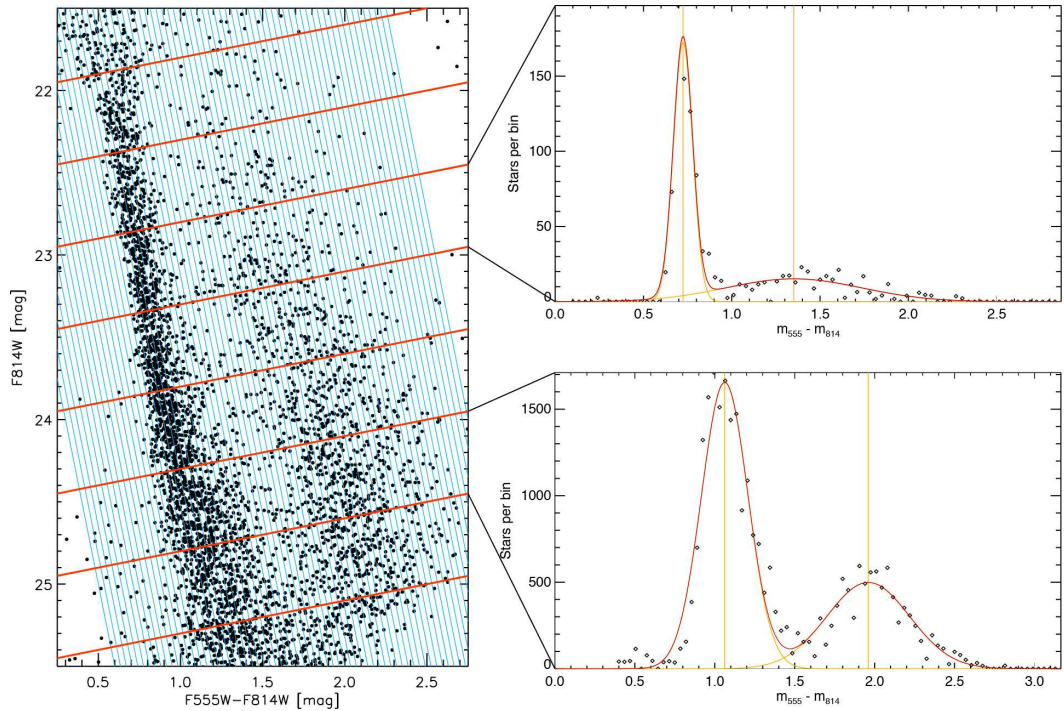}}
\caption{Schematic of the selection of the low-mass PMS stars through 
their number distributions along cross-sections of the faint part of the 
observed CMD. Two examples of these distributions are given, where 
it can be seen that MS and PMS stars are well separated by their 
individual components (peaks) in the distribution. Even at brighter 
magnitudes, where the MS population is most prominent (top distribution), 
the PMS stars can be identified from their second, wider, red peak.
{At fainter magnitudes} (bottom distribution) the distinction between MS
and PMS stars is even more obvious. {The vertical scales in the right hand 
plot are not the same.} \label{f:pmssel}}
\end{figure*}
%%%%%%%%%%%%%%%%%%%%%%%%%%%% FIGURE %%%%%%%%%%%%%

\section{Stellar Content of the region}

\subsection{General description of the stellar content}\label{s:stars}

Previous studies in the region NGC~602/N90  show that it comprises 
three individual dominant stellar clusters and a variety of stellar types \citep{carlson07, 
gouliermis07, schmalzl08, cignoni09, carlson11}. {In particular, apart from
the young cluster NGC~602 located almost at the center of the ring-shaped 
nebula, the observed ACS field covers two more evolved open clusters north of 
the {\sc Hii} ring (see also Fig.~\ref{f:map}). We select the different stellar 
types in the region by first dividing the main-sequence (MS) stars of our 
photometric catalog into Upper MS (UMS) and Low MS (LMS) stars in terms
of their colors and magnitudes in the CMD of  Fig.~\ref{f:cmd}. The UMS stars are 
considered to cover the part of the CMD defined by the conditions: 
$$
\begin{array}{l} 
\displaystyle m_{555}-m_{814} \leq  0.45~~~{\rm and}\\
m_{555}-m_{814} \leq -0.2~m_{814}+4.65. 
\end{array}
$$
This selection is marked in Fig.~\ref{f:cmd} by plotting the  UMS stars
with a light-blue color. The LMS (as well as the low-mass PMS) stars 
are selected to have magnitudes $m_{814} \geq 22$. The remaining part 
of the CMD is considered to be populated mostly by the turn-off and evolved red 
stellar populations. Stars in this part are plotted with grey symbols in Fig.~\ref{f:cmd}.

While UMS stars are the most bright members of all 
three dominant stellar clusters in the region, the LMS stars characterize
{\sl only} the more evolved northern open clusters and the general background
SMC field. This can be seen in the stellar map of the region, shown in Fig.~\ref{f:map},
where stars are again color-coded according to their types. 
There are no low-mass PMS stars in the northern clusters, as they can only 
be found in the central cluster NGC~602 and its immediate surroundings.
This is discussed in more detail in the following section. 
Concerning the covered background field of the SMC, 
it comprises the most evolved observed populations \citep[e.g.,][]{sabbi09}, 
which, apart from the LMS, include the turn-off and the sparse sub- 
and red-giant branches.} There is an apparent overlap between 
the MS turn-off and the PMS-MS transition (turn-on), and therefore these
stars will not be considered in the analysis of the following section to avoid 
confusion by their appearance in the CMD. This exclusion does not affect
the subsequent selection of low-mass PMS stars, because they are selected 
to be fainter than the turn-off magnitude.

\subsection{Low-mass pre--main-sequence stars}\label{s:pmss}

{In the CMD of Fig.~\ref{f:cmd} the low-mass PMS stars 
can be distinguished from the LMS stars, located at the blue 
part of the CMD. Indeed, in previous investigations faint PMS 
stars in the region are selected by their mere location to the 
red part of the CMD. However, the positive identification 
of the true faint PMS stars, especially those closer to the MS, 
requires an accurate quantitative approach in order to 
eliminate any contamination by reddened or binary LMS 
stars. We perform such an analysis by counting the number 
of stars as a function of their ($m_{555}-m_{814}$) colors along 
a series of CMD cross-sections perpendicular to the PMS locus.  
This process is demonstrated in Fig.~\ref{f:pmssel}, where the 
faint CMD of the region is shown (left panel) with the cross-sections 
indicated by the red lines. The cross-sections are selected to have a
width of $\Delta m_{F814W} \sim 0.5$~mag. For the construction of the 
corresponding stellar distribution along each cross-section, stars are 
binned in strips perpendicular to the cross-sections and thus 
approximately parallel to the MS. These strips, which are also almost 
parallel to typical PMS isochrones, are indicated in Fig.~\ref{f:pmssel} 
by the blue lines. }
This  method was originally used for the identification of PMS stars  in $VI$ 
CMDs of Galactic star-forming regions \citep[see, e.g.,][]{sherry04}. Its 
recent application to WFPC2 photometric data of four star-forming regions 
in the Large Magellanic Cloud  showed that {it can be very efficient in 
distinguishing the faint PMS from the MS, even in CMDs with higher confusion 
than the one shown here \citep{gouliermis11}. Two examples of the constructed 
stellar distributions along the CMD cross-sections, i.e., one on a brighter (top) and 
another on a fainter cross-section (bottom), are shown in Fig.~\ref{f:pmssel} 
(right panel). The numbers of stars per bin shown in these distributions, as well
as for all constructed stellar distributions, are corrected for photometric incompleteness 
according to our completeness measurements, performed in \cite{schmalzl08}. }

%\clearpage 
%%%%%%%%%%%%%%%%%%%%%%%%%%%% FIGURE %%%%%%%%%%%%%
\begin{figure}[t!]
\centerline{\includegraphics[angle=0,clip=true,width=\columnwidth]{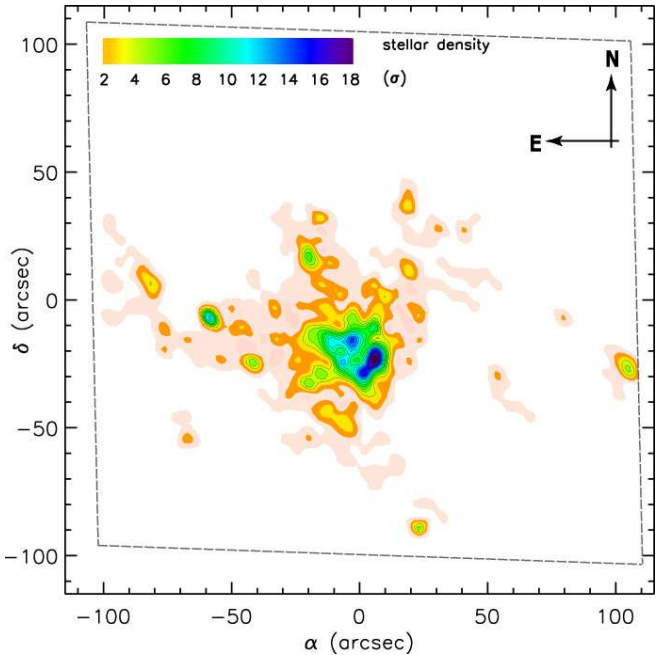}}
\caption{Stellar density contour map of the low-mass PMS stars in the observed region.
Isodensity contours are drawn according to their levels defined in $\sigma$
above the average density, as indicated by the color bar at the top.
This map shows that apart from the main young cluster NGC~602, there
are  distinct compact sub-clusters of PMS stars {located at the inner part and 
at the periphery of the {\sc Hii} ring. The observed ACS field-of-view is indicated by the 
grey dashed lines. Cluster 5 (see also \S~\ref{s:clusters} and Fig.~\ref{f:clusmap}) is at 
the edge of this field.} \label{f:scmap}}
\end{figure}
%%%%%%%%%%%%%%%%%%%%%%%%%%%%%%%%%%%%%%%%%%%%%%

{From Fig.~\ref{f:pmssel} it can be seen that  the distribution 
of stars along the cross-sections through the CMD can be well 
represented by the sum of two distributions; one for the LMS and 
one for the PMS stars. } We fit the number of stars 
per bin of the constructed distributions to a double Gaussian 
function as shown in the examples of Fig.~\ref{f:pmssel}, where
the total fit is plotted by a red line, while the individual components 
by orange lines. The functional form of our fit is \begin{equation} 
\label{eq:gaussfit} y =  \displaystyle 
\sum\limits_{i=1}^{2} N_{i} e^{-0.5((x-\mu_{i})/\sigma_{i})^2}, \end{equation}
where the first term describes the distribution of the LMS  and the
second that of the PMS stars.  We use a least-squares multiple Gaussian fit 
performed by the interactive IDL routine {\sc xgaussfit} (by Don Lindler)
to solve for six parameters, $N_{i}$, $\mu_{i}$, and $\sigma_{i}$, where 
$i=1$ is for the LMS and $i=2$ for the PMS stars. {In this fitting process, 
which is performed for the stellar distributions along each of the selected 
cross-sections, $\mu_{i}$ is the color $m_{555}-m_{814}$ of the peak of 
the  distribution, indicated with yellow vertical lines in the examples of 
Fig.~\ref{f:pmssel}, $N_{i}$ represents the number of stars that correspond 
to the peak of the distribution, and $\sigma_{i}$ defines the width of the 
Gaussian. The full width at half maximum ({FWHM}) of each Gaussian is 
then $2{\sigma_{i}}{\sqrt{2ln(2)}}$ for each cross-section.}
The minimum between the two Gaussian components for every cross-section 
defines the limits between MS and PMS stars on the CMD. {We specified,
thus, the region of the PMS stars on the CMD by a line defined by the minima
where the two Gaussians overlap.} Based on this method we determine the 
 low-mass PMS members in the region. They are color-coded 
in orange in the CMD of Fig.~\ref{f:cmd} and in the stellar map of Fig.~\ref{f:map}.

\section{Spatial distribution of PMS stars in NGC~602/N90}\label{s:clusdet}

The stellar chart of all three dominant stellar types, i.e., UMS, LMS and 
PMS stars, selected as described above, is shown in Fig.~\ref{f:map}.
{There is a definite spatial distinction among
different stellar types, with most of the LMS and UMS stars being concentrated 
in the northern open clusters, while the vast majority of the low-mass PMS, 
along with the hottest (OB-type) stars} in the region as we discuss later, are located at the 
vicinity of the young star cluster NGC~602. This is in line with the findings of previous 
studies, according to which NGC~602 is at an earlier stage of its stellar evolution than the 
northern clusters  \citep[e.g.,][]{schmalzl08}. The fact that the brightest UMS stars are 
indeed located in the vicinity of NGC~602 \citep{hutchings91} signifies further 
its youth. In the chart of  Fig.~\ref{f:map} it can also be seen that 
low-mass PMS stars seem not to be centrally concentrated. There is a clear 
`diffused' distribution of these stars, almost surrounding 
NGC~602, as well as distinct concentrations located within the vicinity of, 
but still away from, the central cluster. We develop an automated algorithm for the
detection of stellar density enhancements and their distinction from
the background density, based on classical star-counts
\citep[see, e.g.,][]{kontizas94}.  The star-count method is found to perform
comparably or usually better than other methods such as {\sl nearest 
neighbor} density or {\sl minimum spanning tree} separation 
\citep{schmeja11}. {We evaluate the spatial distribution of low-mass PMS 
stars in the region quantitatively} and we asses their 
clustering behavior by identifying several ``peripheral'' small compact 
PMS  sub-clusters in the region as statistically important density 
enhancements. We describe this analysis in the  following section.

\subsection{Analysis of the surface density of PMS stars}\label{s:sc}

In order to identify  concentrations of PMS stars
in the observed region, we construct the stellar density map by 
performing star-counts {in a grid of quadrilateral elements}. We
bin the low-mass PMS stars according to their positions 
in a grid of 50~$\times$~50 elements, corresponding to
a scale of about 4\arcsec, i.e., $\sim$~1~pc per grid element.
{This grid element size is empirically chosen after applying several 
tests to grids with elements of different sizes. Smaller sizes of elements in the star-count grid 
allow the appearance of noise in the density maps as 
a large number of small peaks, which correspond to density fluctuations 
rather than true stellar clusterings. On the other hand, grids with larger 
elements would `smear out' individual density peaks by grouping distinct
 compact clusters into larger objects, and thus would not allow us to identify 
 the fine-structure in the surface stellar density of the region. The 
 star-counts grid was aligned according to the stellar chart 
 of Fig.~\ref{f:map}, i.e., it  is oriented so that its x- and y-axes coincide with 
the R.A. and Decl. respectively. The grid  is designed so 
 that the top-left and bottom-right corners of the star-counts map 
 to coincide with those of the observed ACS field-of-view. The derived stellar 
 density map is shown in Fig.~\ref{f:scmap}.}
 In this figure isopleths are drawn in steps  equal to the standard deviation, $\sigma$, of  the 
average density. The lowest  contour corresponds to the 1$\sigma$ density 
level, indicated with faint pink color in Fig.~\ref{f:scmap}.  From this map 
 one can deduce that the large majority of 
low-mass  PMS stars is indeed concentrated in the area of the 
central young cluster  NGC~602, and that the cluster 
itself is {\sl not} symmetric.

%\clearpage 
%%%%%%%%%%%%%%%%%%%%%%%%%%%% FIGURE %%%%%%%%%%%%%
\begin{figure}[t!]
\centerline{\includegraphics[angle=0,clip=true,width=\columnwidth]{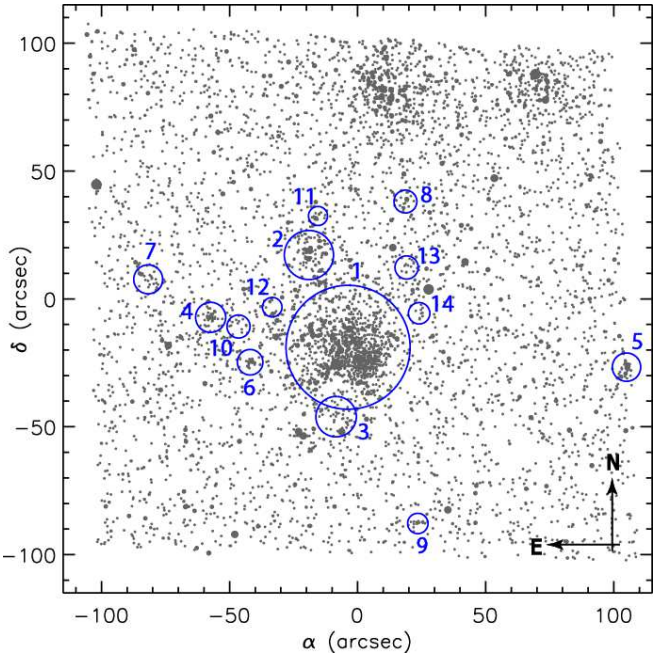}}
\caption{Positions of the PMS clusters identified in the region with star-counts, plotted 
on the full stellar map. The clusters are defined by circular annuli with radii equal to their 
equivalent radii, as given in Table~\ref{t:clus}. They are also annotated by their increscent 
number, as
given in the same table. {All sub-clusters, except 5 and 9, seem to be} contained
in a larger aggregate of stars defined by the average PMS stellar density (Fig.~\ref{f:scmap}). 
This indicates a possible evolutionary and dynamical connection among them.
\label{f:clusmap}}
\end{figure}
%%%%%%%%%%%%%%%%%%%%%%%%%%%%%%%%%%%%%%%%%%%%%%

%\clearpage
%%%%%%%%%%%%%%%%%%%%%%%%%%%%%%%%%%%%%%%%%%%%%%%%%%%%%%%%%%%%
\begin{deluxetable*}{rccccrrrccc}
%\begin{deluxetable}{ccccccc}
\label{t:clus}
\tablecolumns{11}
\tablewidth{0pc}
%\tablenum{1}
\tablecaption{Basic characteristics of the detected PMS star clusters. \label{t:clus} }
\tablehead{
\colhead{Cluster} & 
\colhead{R.A.} &
\colhead{Decl.} &
\colhead{$r_{\rm equiv}$ } &
\colhead{$m_{\rm 555,tot}$} &
\colhead{$N_{\star}$} &
\colhead{$N_{\star,{\rm cc}}$} &
\colhead{$\varrho$} &
\colhead{ \Q\tablenotemark{a}} &
\colhead{ $\sigma_{\Q}$} &
\colhead{YSOs\tablenotemark{b}} \\
\colhead{no.} & 
\multicolumn{2}{c}{(deg J2000)}&
\colhead{(pc)} &
\colhead{(mag)} &
\colhead{} &
\colhead{} &
\colhead{stars~pc$^{-3}$} &
\colhead{} &
\colhead{} &
\colhead{}
} 
\startdata
  1 & 22.383888 & $-$73.560593 & 6.65 & 13.018 & 1028 & 21298&    23.1 & 0.80 &0.02&Y251 \\ 
  2 & 22.397900 & $-$73.550407 & 2.51 & 17.202 &     61 &   1112 &    22.4 & 0.87 & 0.05&Y327 \\
  3 & 22.389244 & $-$73.568047 & 2.02 & 16.194 &     44 &     681 &    26.3 & 0.70 &0.05&Y171 \\
  4 & 22.436224 & $-$73.556877 & 1.45 & 20.830 &     43 &   1390 &  145.2& 0.88 &0.05&Y326 \\
  5 & 22.277424 & $-$73.563522 & 1.35 & 18.636 &     31 &     132 &    17.1 & 0.79 &0.05&Y096\\
  6 & 22.421598 & $-$73.561874 & 1.20 & 19.226 &     25 &     425 &    78.3 & 0.78&0.07&\\
  7 & 22.459705 & $-$73.552544 & 1.39 & 19.246 &     25 &     936 &  111.0 & 0.54&0.08&\\
  8 & 22.360380 & $-$73.544899 & 1.05 & 17.400 &     15 &     395 &  108.7 & 0.69&0.09&\\
  9 & 22.359076 & $-$73.579903 & 0.91 & 23.704 &     14 &     384 &  162.3 & 0.77&0.09&Y090\\ 
10 &	22.425650 & $-$73.557953 & 1.06 & 22.779 &     12 &     532 &  142.3 & 0.85&0.09&\\
11 &	22.393993 & $-$73.546204 & 0.84 & 17.603 &     11 &     534 &  286.9 & 0.72&0.10&\\
12 &	22.412523 & $-$73.555962 & 0.87 & 20.661 &     10 &     322 &  155.7 & 0.62&0.12&Y312\\
13 &	22.360632 & $-$73.552063 & 1.07 & 23.354 &       9 &     267 &    69.4 & 0.94&0.12&\\
14 &	22.356310 & $-$73.557083 & 0.97 & 22.750 &       8 &     128 &    44.7 & 0.91&0.12&\\
\enddata
\tablenotetext{1}{The derivation of the parameter \Q\ and typical uncertainties $\sigma_{\Q}$ in its measurement are described in \S~\ref{s:mst}.}
\tablenotetext{2}{YSO candidates IDs from \cite{carlson11}.}
\tablecomments{For explanations on the columns see \S~\ref{s:clusters}. {The total brightness $m_{\rm 555,tot}$ 
of each cluster is the cumulative brightness of all stars identified within the boundaries of the cluster. $N_{\star}$ is the 
total number of observed stars per cluster, and $N_{\star,{\rm cc}}$ is the same number with completeness corrections 
applied.}  Sub-cluster~1 corresponds to the main cluster NGC~602.}
\end{deluxetable*}
%%%%%%%%%%%%%%%%%%%%%%%%%%%%%%%%%%%%%%%%%%%%%%%%%%%%%%%%%%%

In addition, the density map of Fig.~\ref{f:scmap} shows that 
apart from the central cluster, there are several `clumps' 
of PMS stars, which can be seen within the boundaries of the 
lowest density isopleth.  These are small in size compact sub-clusters
of PMS stars, which stand out due to their high stellar density. 
Outside these clusters {a diffuse PMS 
stellar distribution fills the space between them.}
Both these features in the surface density of low-mass PMS
stars  reveal their clustering behavior in the region NGC~602/N90. 
This behavior can be described as most of the PMS stars being 
centrally concentrated in the cluster NGC~602 and in small compact 
peripheral sub-clusters, {which in turn seem to be surrounded by a general 
diffuse distribution of the same type of stars.}

\subsection{Identification of PMS Sub-Clusters}\label{s:clusters}

According 
to our methodology, we consider any stellar concentration 
revealed in the map of Fig.~\ref{f:scmap}\ by the isopleths
that correspond to stellar density of at least 3$\sigma$ above 
the background (indicated by the yellow-colored contours) 
as being {\sl statistically important}, and thus as a candidate star 
cluster. As discussed above,  these clusters are not completely 
isolated from each other, but they appear to be members 
of a larger concentration of PMS stars. We 
define the boundaries of each of these 
clusters by their limits according to the 2$\sigma$ isopleth
in the density map, and we measure the 
corresponding `equivalent' radius as \beq r_{\rm equiv} = \sqrt{\frac{A}{\pi}}~~,\eeq 
where $A$ is the surface enclosed by the threshold  
isopleth, assuming circular symmetry
in their shapes. We select the stellar members of each 
cluster from the original photometric catalog of all stars
(not PMS stars alone), by the positions of the stars as
they are confined by the 2$\sigma$ isopleth as well.
With this method we identified 14 distinct PMS sub-clusters,
NGC~602 included. 

%\clearpage 
%%%%%%%%%%%%%%%%%%%%%%%%%%%% FIGURE %%%%%%%%%%%%%
\begin{figure*}[t!]
\centerline{\includegraphics[angle=0,clip=true,width=0.725\textwidth]{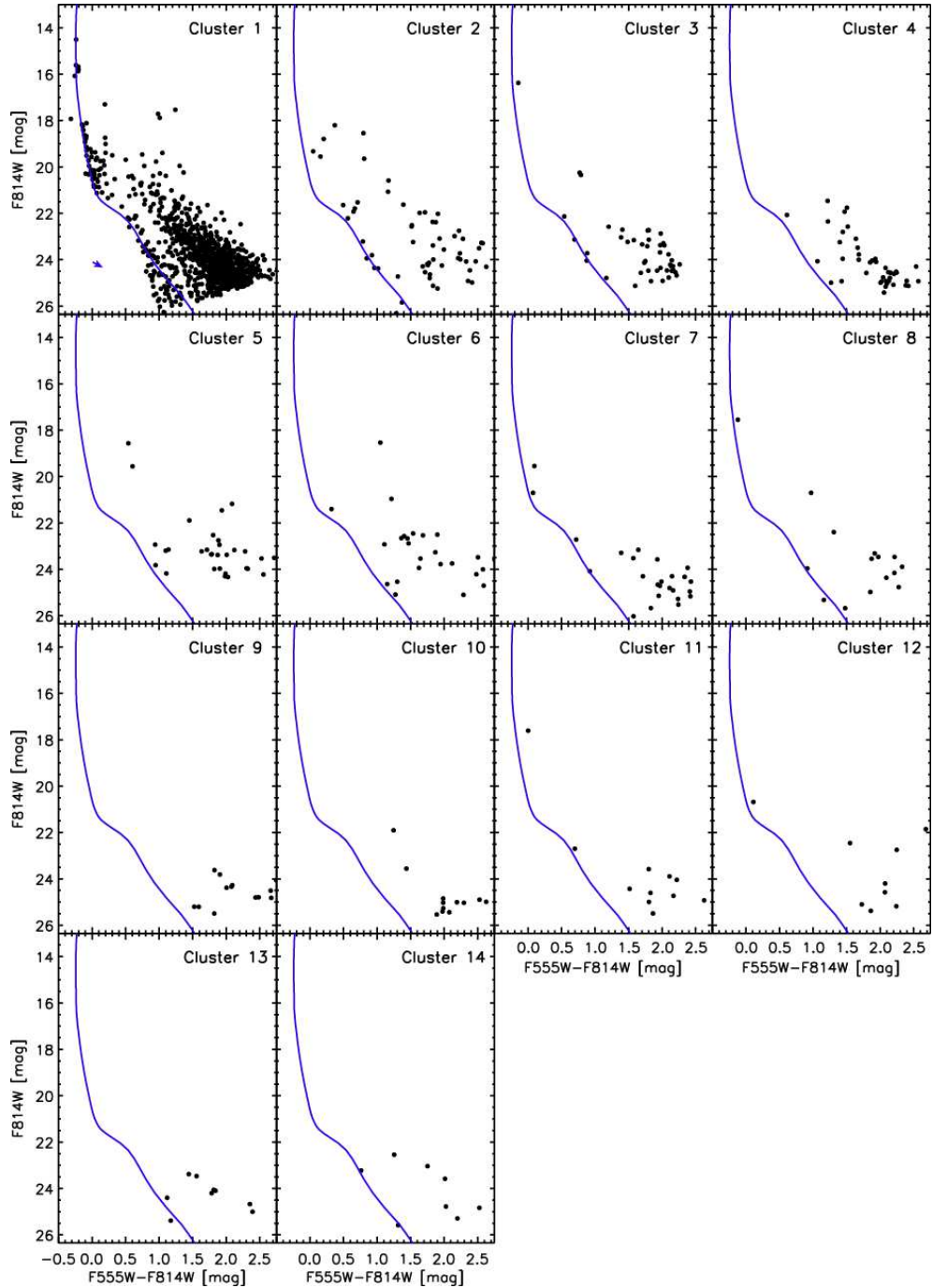}}
\caption{F555W$-$F814W, F814W CMDs of the PMS sub-clusters found in the 
region NGC~602/N90. The ZAMS from the Padua set of models \citep{girardi02}
{for the metallicity of the SMC ($Z=0.003$)}, is plotted in blue for guidance. It is corrected for a distance modulus of 
$m-M \simeq$~18.95~mag, and an extinction of $A_{V} \simeq$~0.3~mag
\citep{schmalzl08} represented by the reddening vector {plotted at the 
lower left  quadrant of Cluster 1 CMD}.\label{f:cluscmds}}
\end{figure*}
%%%%%%%%%%%%%%%%%%%%%%%%%%%%%%%%%%%%%%%%%%%%%%

The positions of the identified sub-clusters are annotated in the stellar 
map of Fig.~\ref{f:clusmap}, and their characteristics
are given in Table~\ref{t:clus}. The identity numbers of the sub-clusters 
are given in Col. 1 of the table. Their coordinates, given in Cols. 2 and 3, 
are derived from their baricenters, i.e., the average positions
 of all their stellar  members.  Their equivalent 
 radii and total brightness (not corrected for incompleteness) 
 are given in Cols. 4 and 5 respectively. The original total numbers
 of stars (of all types)  in each 
 sub-cluster are shown in Col. 6 of Table~\ref{t:clus}, and in Col. 7 these
 numbers are corrected for incompleteness according to our 
 completeness measurements in \cite{schmalzl08}. Naturally, since 
 we are limited by the detection efficiency of our photometry,
 these corrections are limited to $\sim$~50\% completeness 
 and therefore they do not reflect the true total stellar numbers
 of each sub-cluster. They can be used though for an estimate of the
 volume stellar density of the clusters, which reflects their 
 compactness. We give this estimate for every sub-cluster in Col. 8
 of the table. The \Q\ parameter, {which permits us to quantify} the
structure of a cluster and to distinguish between 
centrally concentrated and hierarchical clusters and 
typical uncertainties, $\sigma_{\Q}$, in its determination 
(see \S~\ref{s:mst}) are given in Cols.~9 and 10 respectively. 
Finally, the YSO candidates  from the catalog of 
 \cite{carlson11}, which are found to match the positions
 of our sub-clusters (see \S~\ref{s:clusstars}) are given in Col. 11 of 
 Table~\ref{t:clus}. {It should be noted that among the identified 
 sub-clusters, cluster~5 being located on the edge of the observed 
 field-of-view, extends somewhat out of this field as can be seen in 
 IRAC images of the region \citep[see, e.g.,][]{carlson11}. Therefore, its derived parameters should
 be considered as the lower limits.}

\subsection{Stellar content of the identified clusters}\label{s:clusstars}

The individual CMDs of the detected PMS clusters are shown 
in Fig.~\ref{f:cluscmds}. These CMDs are constructed from all
stars included within the boundaries of the sub-clusters,
as described above, their numbers being given in Col.  6 of 
Table~\ref{t:clus}. As shown in this figure,  these 
numbers for about half of the sub-clusters are very low, not enough
to extract significant information. We show the 
CMDs with an overlaid {\sl Zero Age Main Sequence} (ZAMS) 
for guidance. The CMDs of the 
less populous sub-clusters, with identification numbers 9 
and higher, are shown in the third and forth panels of 
Fig.~\ref{f:cluscmds}. It is interesting that while some 
sub-clusters host stars in a wide range of magnitudes, other and in 
particular the less populous host mostly only faint PMS stars, 
down to our detection limit. It is also worth noting that 
some of  the detected sub-clusters appear bright in mid-IR 
wavelengths as seen with the {\em Spitzer Space Telescope} (STT).  
In particular, we matched 
the positions of our sub-clusters with those of the candidate YSOs 
population in the region derived by \cite{carlson11},
and we found that seven of the sub-clusters coincide with one YSO each; we provide 
the YSO identities in Col. 11 of Table~\ref{t:clus}. It should be noted that
YSO candidate Y251 does not quite coincide with the cluster
NGC~602, {but is on the eastern edge within its boundaries. 
The other YSO candidates coincide with the whole extent of their 
matching star clusters, which are very small.}
This is a natural result of the low resolving power of SST. 

Low-mass PMS stars in star-forming regions of the MCs 
are widely spread in the red-faint part of their optical 
CMDs \citep[see, e.g.,][for recent results]{cignoni09, vallenari10, 
gouliermis11}. As a consequence, according to PMS evolutionary 
models the CMD positions of these stars imply a broad coverage 
in ages, which may be indicative of an age-spread. However, 
previous investigations have shown that the broadening of the 
 loci of faint PMS stars in the optical CMD does not 
 necessarily provide proof of an age-spread, since it may well be the result 
 of biases introduced by observational constrains,  i.e., photometric accuracy 
 and confusion. The CMD positions of these stars depend also on their 
 physical characteristics, such as variability, binarity and circumstellar 
 extinction (see, e.g., \citealp{dario10} and \citealp{ jeffries11}, for 
 detailed discussions), making the determination of their ages from
 isochrone fitting quite challenging. Under these circumstances, it is 
 not possible to extract absolute ages of such stars, 
 unless thorough modeling of their positions is being applied. 

In star-forming regions differential reddening may also 
be considered as an important factor for the 
broadening of faint PMS stars in the observed CMDs, but in 
NGC~602/N90, according to our measurement of 
$\langle A_{V} \rangle \simeq 0.2$ with a variance of $\simeq 0.07$ 
\citep{schmalzl08}, interstellar reddening 
cannot  be accounted for the observed CMD-broadening. A reddening 
vector corresponding to $A_{V} \simeq 0.3$ is shown 
in the top-left corner of the first CMD of Fig.~\ref{f:cluscmds} 
{to demonstrate the small effect of variable optical extinction to the observed CMD 
positions of the PMS stars. As we discuss later, some of the
identified sub-clusters show bright IR emission, coinciding with 
known YSOs \citep{carlson11}, indicating that possibly they are 
still embedded in their natal clumps.} Considering the above discussion about the
CMD-broadening of PMS stars in all detected sub-clusters, we use here a set 
of PMS evolutionary models not to 
determine their ages from their PMS populations, but to define an 
upper limit for these ages on  a comparative basis.
It is worth noting that the stellar numbers included in each sub-cluster are too few
to provide a solid age determination, except of very few cases. 

%\clearpage 
%%%%%%%%%%%%%%%%%%%%%%%%%%%% FIGURE %%%%%%%%%%%%%
\begin{figure*}[t!]
\centerline{\includegraphics[angle=0,clip=true,width=0.875\textwidth]{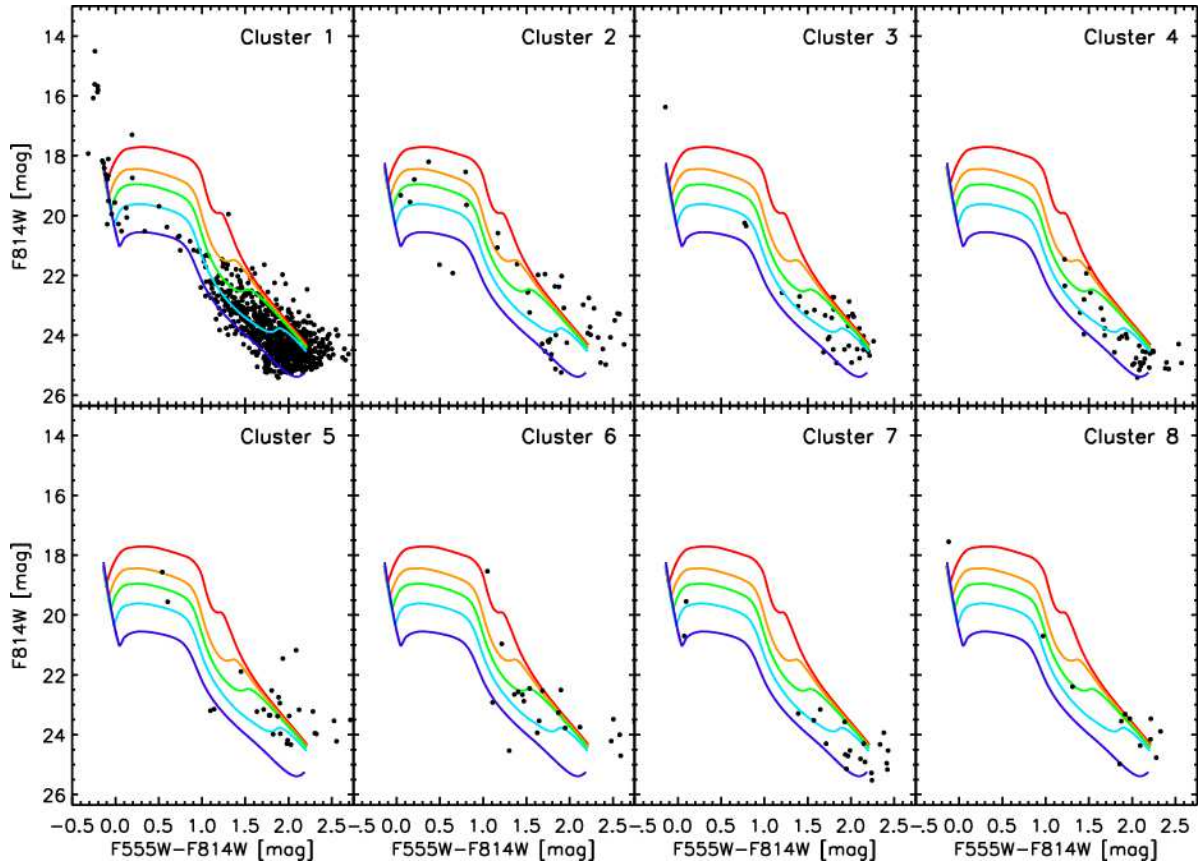}}
\caption{F555W$-$F814W, F814W CMDs of the eight most populous PMS sub-clusters
in the region NGC~602/N90. PMS isochrone models calculated with {\sc franec} 
\citep{tognelli11} are overlaid in different colors to demonstrate that while an absolute
age for the sub-clusters is not possible to be determined, an upper limit for this age
can be well established. The models correspond to ages of 0.5 (red), 1.0 (orange), 
1.5 (green), 2.5 (light blue), and 5.0 (blue) Myr. They are corrected for distance 
and extinction.\label{f:cmdpmsiso}}
\end{figure*}
%%%%%%%%%%%%%%%%%%%%%%%%%%%%%%%%%%%%%%%%%%%%%%

We used the most recent Pisa PMS models which rely on the state-of-the-art
of input physics and are available for a wide range of $Z$, $Y$, $\alpha$,
mass, and age values\footnote{The Database is available at\\
\url{http://astro.df.unipi.it/stellar-models/} } 
\citep[see][for a detailed description]{tognelli11}.
These models have been computed with an updated version of the 
{\sl Frascati Raphson Newton Evolutionary Code} ({\sc franec}), 
a well-tested Henyey evolutionary code \citep{deglinnocenti08},
for the canonical metal abundance of the SMC, namely [Fe/H] = $-$0.65 ($Z=0.003$
and $Y=0.254$). The Pisa PMS database provides models with three different
 values of the mixing-length parameter $\alpha$, namely 1.2, 1.68 (solar calibrated)
 and 1.9. In this study we use tracks with $\alpha = 1.2$, which provide
the best fit of the data in agreement with other indications from binary systems 
\citep[][]{simon00, steffen01, stassun04, gennaro11} and Li-depletion 
\citep[][]{ventura98, dantona03, tognelli11b}  studies. We transformed tracks and isochrones from the theoretical plane
into the ACS/WFC VEGAmag photometric system for a direct comparison with our
photometry by using the synthetic spectra provided by
\cite{brott05} for $T_{\rm eff} \le 10000$ K
and by \cite{castelli03} for $T_{\rm eff} > 10000$ K.

In Fig.~\ref{f:cmdpmsiso} we
show the CMDs of the eight most populous sub-clusters in our sample, 
i.e., clusters that include enough stars for a meaningful comparison with 
isochrones. {The contribution of the field stellar population is removed 
from the CMDs using a statistical subtraction technique based on 
the Monte Carlo method} after selecting the least populated area of the 
region as the most representative of the field population \citep[see][for a 
detailed description]{schmalzl08}. Indeed, a comparison between the CMDs of 
Fig.~\ref{f:cmdpmsiso} with those of Fig.~\ref{f:cluscmds} for the same
clusters show that the LMS, as well as the turn-off and sub-giants 
are essentially 
eliminated after the decontamination of the field. This subtraction is 
in particular interesting for  the most populous cluster, i.e., 
NGC~602 (cluster~1). In Fig.~\ref{f:cmdpmsiso}  {\sc franec} 
PMS isochrones for ages 0.5, 1, 1.5, 2.5 and 5~Myr are overlaid
with different colors. From these isochrones one can deduce that all  
sub-clusters are quite young, with ages \lsim~5~Myr, considering the broadening 
of the PMS stars and the fact that the 5~Myr model  
is quite away to the blue from the observed CMD-location of the faint PMS 
populations. Nevertheless, from the inspection of the individual CMDs 
one can also conclude that not all sub-clusters appear to have the same age, {with 
clusters 2, 5 and 6 appearing to be as young as 1~Myr. }

While the observed broadening of the CMD-positions of faint PMS stars with 
$m_{\rm F814}$~\gsim~21 does not allow us an accurate age determination for the sub-clusters, the 
CMD-locations of the brighter PMS stars are less affected by photometric 
uncertainties or other sources of luminosity spread, such as variability, unresolved 
binarity or circumstellar extinction and therefore they are more successfully fitted 
by the appropriate isochrones.  Moreover, considering that these stars are located 
at the turn-on in the CMD, where the isochrones of different ages are more distinct 
to each other, these stars are more accurate chronometers of their hosting clusters.
We consider, thus, the bright PMS stars of the CMDs shown in Fig.~\ref{f:cmdpmsiso} 
with magnitudes 18~\lsim~$m_{\rm 814}$~\lsim~21 and colors 0.0~\lsim~$m_{\rm 555} 
- m_{\rm 814}$~\lsim~1.5. A comparison of the CMD-locations of these stars in respect 
to the overlaid isochrones between different sub-clusters shows that indeed 
it seems that they do have different ages.

In particular, the bright PMS stars in both the most well-populated 
sub-clusters 1 and 2 seem to fit very well the PMS-MS transit phase, i.e., 
the turn-on, but for different ages. Specifically, taking the isochrones fit  
of these stars at `face value' it seems that sub-cluster~1 (NGC~602) 
is older than sub-cluster 2, since the turn-on of sub-cluster~1
fits better to an age of \lsim~5~Myr, while that of sub-cluster~2 
to an age of \lsim~2.5 Myr. As far as the rest of the sub-clusters are 
concerned, the very few bright PMS stars in them do
not allow definite age distinctions from the others. However, 
the positions of these stars, where are apparent, show a 
trend to younger ages and in particular around 2.5~Myr. 
This result  is in agreement with that of \cite{cignoni09}, 
who found that star formation in the region NGC~602/N90 
in the recent 10 Myr has been quite high, reaching a peak in the last 
2.5~Myr. Nevertheless, it can only serve 
as an indication of a possible age difference between the sub-clusters, 
and not as a direct proof. 

No additional information can be provided 
by the faint PMS populations of the sub-clusters. From these populations,
and the comparison of their CMD positions with the PMS isochrones 
we see that in some sub-clusters (e.g., sub-clusters 2, 5, and 6) faint PMS stars 
are extremely  red,  located outside the coverage of the models, while 
in others (e.g., sub-clusters 3 and 4) they appear well distributed among  
the selected models. {This difference may reflect age differences among 
the clusters, in the sense that redder stars are younger.  Higher extinction in 
the most embedded clusters can also shift stars redward on the CMD, making 
them appear younger.}

Concerning the candidate YSOs identified in the sub-clusters 
\citep[][see also Table~\ref{t:clus}]{carlson11}, as mentioned 
earlier there is one candidate included within the outer boundaries
of sub-cluster 1, but not quite coinciding with the main part of the cluster. 
On the other hand, each of sub-clusters~2, 3, 4 and 5 does coincide 
at its complete extent with IR-bright candidate massive YSOs. 
Previous independent studies that combine observations from both
HST and SST note that objects that appear as massive YSOs in 
mid-IR bands can be resolved into multiple PMS sources in the 
optical \citep{gouliermis07, carlson07, carlson11}. Here, we 
identify some of them as compact PMS sub-clusters that host 
objects at very early evolutionary  stages, indicating their 
extreme youthfulness. Sub-cluster~2 is a characteristic 
example of a compact cluster at earlier stages of formation, since 
it is highly embedded \citep[see, e.g.,][]{carlson07, gouliermis07}, as 
opposed to sub-cluster~1, which appears to be clean from any nebula.  
Candidate YSOs are found to coincide with two additional sub-clusters
(9 and 12), for which the few faint PMS stars cannot give us
a reasonable age estimation. The bright IR emission of these
sub-clusters, as well as their few optically detected PMS members 
indicate that these are highly embedded clusters.

%\clearpage 
%%%%%%%%%%%%%%%%%%%%%%%%%%%% FIGURE %%%%%%%%%%%%%
\begin{figure*}[t!]
\centerline{\includegraphics[angle=0,clip=true,width=\textwidth]{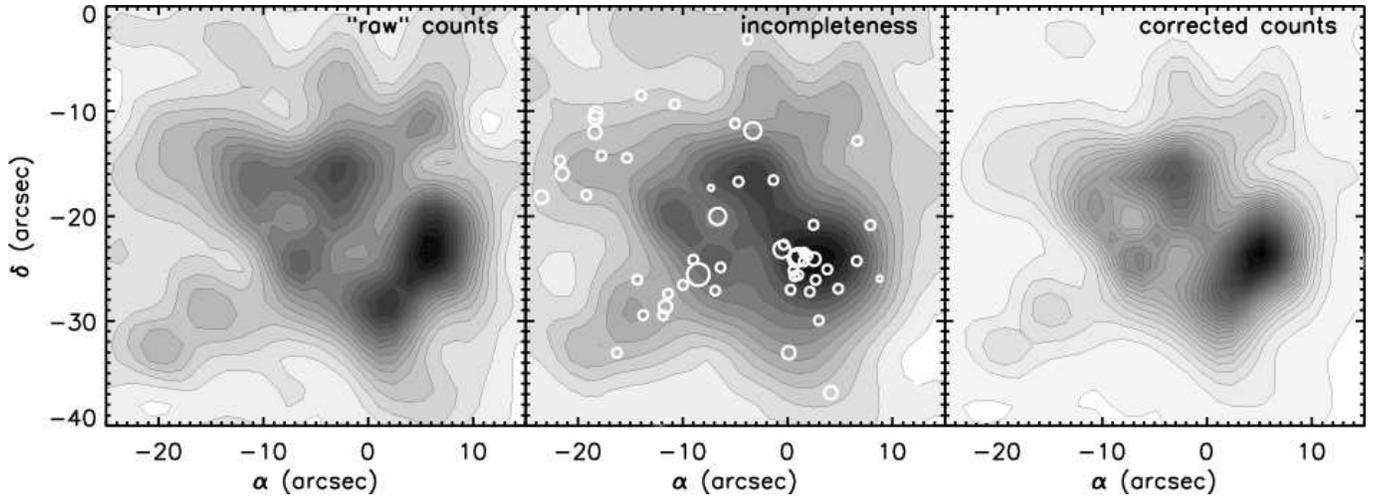}}
\caption{Isodensity contour map of the central cluster NGC~602. {\em Left}: The map constructed 
from `raw' counts on the original observed numbers of low-mass PMS stars with no correction 
for incompleteness applied. {\em Middle}: The corresponding `completeness' map, constructed 
by counting stars from our fake star experiments. Dark grey contours correspond to areas with the 
least success in recovering stars in the photometry, i.e., with the higher incompleteness. Light 
grey contours correspond to areas with the highest detecting efficiency, i.e., higher completeness 
in our photometry for low-mass PMS stars. Open white circles correspond to the UMS in the area
of NGC~602. There is an apparent relation between incompleteness and high concentration of
such stars, indicating that crowding affects our photometric detection efficiency. {\em Right}: Same 
as the map in the left panel, but corrected for incompleteness according to the measurements
shown in the middle panel. This density map shows that sub-clustering within the boundaries
of the cluster is not the result of incompleteness alone, but a real phenomenon. {All maps are
plotted at the same scale.}
\label{f:cclcccm}}
\end{figure*}
%%%%%%%%%%%%%%%%%%%%%%%%%%%% FIGURE %%%%%%%%%%%%%

\subsection{The Distribution of PMS stars within NGC 602}

The main cluster of the region, NGC~602, is the dominant PMS stellar 
concentration. While 
NGC~602 is being classified as a single group, based on the  
2$\sigma$ isopleth in the stellar density map of Fig.~\ref{f:scmap}, one can 
observe in the same map that there are multiple high density peaks within 
the boundaries of the cluster, suggesting that the cluster itself is a 
multiple stellar group. However, there are two observational biases,
which can affect the two-dimensional spatial distribution of stars in a 
cluster, giving 
false evidence of  multiple stellar density peaks. The first is extinction, 
which hides the 
fainter stars of the cluster with differential reddening within the cluster 
allowing only patches of a homogeneous stellar concentration to 
appear as density peaks. The second is confusion in the stellar photometry, 
which due to crowding or high sky contamination can produce an effect 
of multiplicity, since the inefficient detectability of stars at some parts of the cluster would 
produce the appearance of  different peaks within the cluster.  
Concerning the first bias a maximum value 
for extinction of $A_{V} \simeq$~0.3~mag, based on our reddening measurements 
\citep{schmalzl08} is not high enough 
to significantly hide portions of stars in the cluster. The low extinction 
of the cluster is further verified by the ancillary H{\alp} HST observations 
and from imaging with STT of the region \citep[see, e.g., Figs.~1 and 2 respectively 
in][]{carlson07}. These observations show that the gas and dust content in the 
bubble of the region is quite low, being removed towards 
its edges and leaving the cluster quite clean from extinction.

On the other
hand, confusion in the detection of stars, and therefore incompleteness 
in our photometry, may be a more important factor in the appearance of 
multiple density peaks in the cluster. We quantified the incompleteness
in our photometry on the basis of artificial star experiments in \cite{schmalzl08}.
As a consequence, we are
able to determine the completeness of our photometry not only 
as a function of  magnitude and color, but also of position. 
In order to assess if indeed incompleteness is enough to explain 
the apparent multi-peaks in stellar density within NGC~602, we construct
an {\sl incompleteness map} of the area of the cluster, 
{by applying the same star-counts technique to the artificial stars catalog
with that we used for the 
construction of the stellar density map of Fig.~\ref{f:scmap}.} For reasons 
of comparison, we construct the incompleteness map with the same input  
parameters as for the stellar density map of the region
(see \S~\ref{s:sc}). 

In Fig.~\ref{f:cclcccm}~(left panel) we show the part
of the stellar density map of Fig.~\ref{f:scmap} centered on NGC~602.
In the middle panel of the figure is shown the incompleteness map 
for the same area. Darker grey tones correspond to higher incompleteness,
i.e., lower detectability of stars. We overlay the positions of the most bright 
UMS stars in the area of the cluster to demonstrate that higher incompleteness 
is related to these stars, and thus indeed it is the product of confusion. 
{We do not count stars with completeness less than 40\% in both 
filters, as derived with our artificial star experiments in \cite{schmalzl08}.} 
We correct the original star-count map of the cluster according to the 
incompleteness map. From the completeness-corrected map, shown in the right 
panel of Fig.~\ref{f:cclcccm}, it is seen that the south-western double density 
peak, at the right of the original map (left panel), has become a 
larger single peak. However, the appearance of different peaks within the cluster has
not been eliminated, even after our correction for incompleteness. This gives evidence 
that the fragmented appearance of the cluster is real, and that {photometric confusion can only partly account for it.}

\subsection{The diffuse PMS population}\label{s:diffpop}

 {The spatial distribution of low-mass PMS stars, shown
in Fig.~\ref{f:scmap}, demonstrates the existence of statistically 
significant distinct stellar concentrations, which we 
classify as compact PMS sub-clusters in the previous 
sections. However, the spatial distribution of low-mass PMS stars 
shows also that these sub-clusters are surrounded by 
a general diffuse distribution of the same kind of stars. This clearly 
indicates that while these sub-clusters are distinct stellar groups, 
they are not totally independent. They are possibly dynamically and even 
evolutionarily related to each other.} The diffuse
PMS population accounts for the $\sim$~40\% of the total PMS
population observed in the region, i.e., 60\% of the total detected 
PMS stars are clustered. This fraction is higher than in the SMC star-forming 
region NGC~346/N66, where we found only about 40\% of the PMS stars in 
clusters \citep{schmeja09}.

\begin{figure}[t!]
\centerline{\includegraphics[angle=0,clip=true,width=0.875\columnwidth]{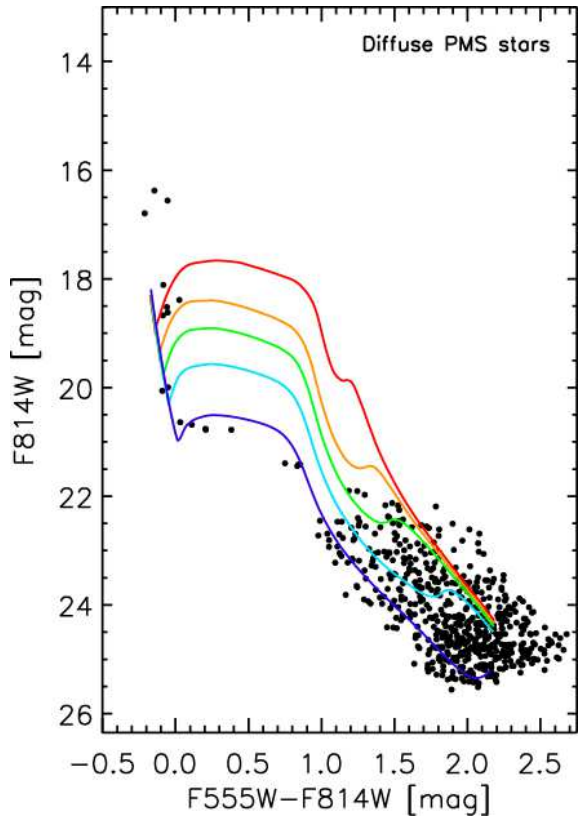}}
\caption{F555W$-$F814W, F814W CMD of the diffuse PMS stars in the 
region NGC~602/N90. PMS isochrone models calculated with {\sc franec} 
\citep{tognelli11} are overlaid as in Fig.~\ref{f:cmdpmsiso}. They demonstrate
that PMS stars not hosted by clusters possibly share the same formation history
with those in NGC~602, being somewhat older than the members 
of the compact PMS sub-clusters.  \label{f:diffisocmd}}
\end{figure}

We consider as members of the 
diffuse population all PMS stars located at areas of stellar density 
less than 2$\sigma$, as indicated by the pink color in the density 
map of Fig.~\ref{f:scmap}. This selection is based on the fact 
that the 2$\sigma$ isopleth sets the boundaries of all detected sub-clusters. 
The field-subtracted CMD of the diffuse PMS stars is shown in 
Fig.~\ref{f:diffisocmd} with a set of {\sc franec} PMS 
isochrones similar to that shown in Fig.~\ref{f:cmdpmsiso} overlaid.
This CMD includes about 500 faint PMS stars, the CMD-loci of which
demonstrate the same broadening as in the CMDs of the clusters, and
therefore any age determination from them would not be accurate. On the 
other hand the few bright PMS, located on the transition between 
PMS-MS seem to fit best the 5~Myr isochrone, which clearly indicates 
that these stars may be somewhat older than the members of the 
small compact PMS sub-clusters, having ages close to those of NGC~602.

%\clearpage 
%%%%%%%%%%%%%%%%%%%%%%%%%%%% FIGURE %%%%%%%%%%%%%
%%%%%%%%%%%%%%%%%%%%%%%%%%%%%%%%%%%%%%%%%%%%%%

\section{Minimum Spanning Tree and the \Q\ parameter}\label{s:mst}

In order to further quantify the clustering of low-mass PMS stars
in the region NGC~602/N90, we apply a second method, which 
makes use of a minimum spanning tree (MST), defined as the unique 
set of straight lines (`edges') connecting a given set of points without 
closed loops, such that the sum of the edge lengths is a minimum 
\citep{boruvka26, kruskal56, prim57}. In particular we make use of 
the  \Q\ parameter \citep{cw04}, defined as $\Q = \bar{\ell}_{\rm MST}/\bar{s}$, 
which combines the normalized correlation length $\bar{s}$,
i.e.\ the mean distance between all stars, and the normalized mean edge 
length $\bar{\ell}_{\rm MST}$ derived by the MST for the cluster.
It is a measure of the fractal dimension $D$ of a stellar group, {permitting 
us to quantify} the structure of a cluster and to 
distinguish between clusters with a central density concentration and 
hierarchical clusters with possible fractal substructure.
Large \Q\ values ($\Q > 0.8$) describe centrally condensed clusters
having a volume density $n(r) \propto r^{-\beta}$, while small \Q\ values 
($\Q < 0.8$) indicate clusters with fractal substructure. {This method and the
interpretation of \Q, as developed by \cite{cw04}, is actually based on
the three-dimensional structure of stellar systems. In particular, these 
authors constructed artificial three-dimensional clusters and studied the 
\Q\ values for different two-dimensional projections to derive the correlation 
of \Q\ with the radial density exponent $\beta$ or the fractal dimension $D$.} 
\Q\ is correlated with $\beta$ for $\Q > 0.8$ and anticorrelated with $D$ for 
$\Q < 0.8$. {\cite{cw04} show that unlike centrally concentrated clusters, 
fractal clusters can look quite different in different projections, leading to a 
wider range in the \Q\ parameter values. As a consequence, the three-dimensional
structure affects more the fractal systems and this is the reason why the derived 
errors are larger for low \Q\ values.} A detailed description of the 
method, and in particular its implementation in this study, is given by 
\cite{sk06}.

We assess the structural status of the PMS sub-clusters detected in the region 
NGC~602/N90 by applying the MST method on the samples of PMS stars of each 
sub-cluster. One important limitation to this application is the total number of points 
in each set. With decreasing number of objects, $n$, the error of \Q\ increases, {and 
the value  becomes less meaningful.} We made an estimation of the \Q\ measuring 
accuracy with the use of artificial clusters, which we constructed with specific density 
profiles, and for which we computed the average \Q\ and its standard  deviation, 
$\sigma_{\Q}$, from 100 different realizations. From our experiments we found {a 
general degeneration of \Q\ with decreasing $n$.} {For smaller $n$ the value of \Q\ 
for centrally concentrated clusters stays roughly constant and higher than 0.8. On the 
other hand, for fractal clusters with smaller $n$, \Q\ rises toward and even over 0.8, 
providing false evidence that they are centrally concentrated.}

This behavior was more apparent for artificial clusters with \lsim~30 members.
Therefore, while in Table~\ref{t:clus} we give a \Q\ value for all sub-clusters 
detected in NGC~602/N90, our analysis will be mainly based only to those with 
more than 30 members (where $\sigma_\Q \lesssim 10\%$), i.e., for the five more 
populous sub-clusters. The derived \Q\ values and typical uncertainties in their 
determination, derived from our artificial clusters experiments, are given in Cols.~9 
and 10 of Table~\ref{t:clus} respectively. From these values it can be seen that almost 
all considered clusters lie on the limit between being centrally concentrated and fractal. 
In particular NGC~602, which is quite populous and therefore its \Q\ is more accurately 
determined, has the intermediate value of $\Q = 0.8$. Among the sub-clusters with more 
than 30 members only one, sub-cluster~3, has a \Q\ somewhat smaller but still very close 
to this threshold. About two-thirds of the poorer sub-clusters, i.e., those with less than 30 
members, have \Q\ values that classify them as  centrally concentrated, or position them 
very close to the limit between centrally concentrated and fractal clusters. However, as 
mentioned above, this result is unreliable, because of the lower accuracy in the 
determination of the \Q\ parameter of these clusters. It should be noted that all \Q\ 
values are derived from the faint PMS stars alone, but the use of both PMS and UMS 
stars included in each sub-cluster delivers almost the same values, not altering at all 
our results, {possibly because the number of additional sources is very low.}

%\clearpage 
%%%%%%%%%%%%%%%%%%%%%%%%%%%% FIGURE %%%%%%%%%%%%%
\begin{figure}[t!]
\centerline{\includegraphics[angle=0,clip=true,width=\columnwidth]{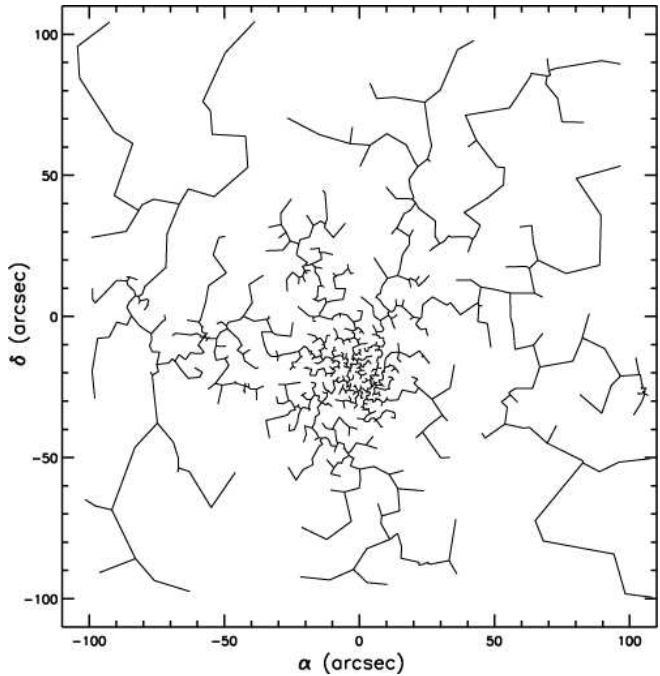}}
\caption{The MST graph for the entire low-mass PMS population observed in the region NGC~602/N90. 
\label{f:mst}}
\end{figure}
%%%%%%%%%%%%%%%%%%%%%%%%%%%% FIGURE %%%%%%%%%%%%%

We assess the structural
behavior of the diffuse PMS stars by considering the sample of all PMS stars,  
the density of which corresponds to the average stellar density in the region, 
as if the whole structure represents one stellar concentration.
The reason for this treatment, which also considers  all detected sub-clusters 
 as part of the diffuse population, is that there would be no physical 
meaning to determine the behavior only of the PMS stars in the `inter-cluster' area,
since such a patchy stellar distribution is not real. {This application of the MST
method for the complete sample of PMS stars derives  a \Q\ value of  0.85, well
above the threshold of 0.8 toward the \Q\ values for centrally concentrated clusters.
This result provides solid evidence that the whole agglomeration of PMS stars 
in the region can be itself treated as a centrally concentrated 
cluster as well (see also Fig.~\ref{f:mst}).} However, 
{if  the UMS stars are considered also} the value drops to $\Q = 0.75$, due to the more 
`clumpy' distribution of these stars. While this value is still not low enough to safely 
characterize the structure as fractal, it certainly suggests that this concentration is 
hierarchically structured. In any case, our results indicate that
the diffuse PMS population is dynamically related, {if not bound}, to the clustered
population, and that the individual PMS sub-clusters themselves are tightly related to
each other. 

\section{A scenario of clustered star formation in NGC~602/N90}\label{s:scenario}

Our results above on the clustering behavior of PMS stars, in conjunction to those 
on the age differences between the observed PMS populations in the individual sub-clusters and 
the inter-cluster area (\S\S~\ref{s:clusstars} and \ref{s:diffpop}), leads us to the construction of 
a star cluster formation scenario in NGC~602/N90. According to this scenario 
PMS stars were formed \lsim~5~Myr ago clustered at the center of the bubble, where
NGC~602 is located, with a swift following formation of stars $\lsim$~2.5~Myr ago in the peripheral small 
compact sub-clusters. {It is worth noting that these timescales are  consistent with those 
derived from the YSOs in the region by \cite{carlson11}.}
This scenario is also in a very good agreement with the result by \cite{nigra08}
that star formation in the region initiated $\sim$~4~Myr ago, and with the results by \cite{cignoni09}, 
according to which the star formation rate in the region seems to have increased with time, 
with a peak at $\sim$~2.5~Myr ago. This short timescale between the two indicative stellar generations 
is in accordance to the paradigm of rapid star formation on dynamical time-scales \citep[e.g.,][]{elmegreen00, 
hartmann01}, as opposed to that of slow star formation \citep[e.g.,][]{shu87}. 

The short timescale between the star formation events may also explain the origin of the diffuse PMS stars. 
Theoretical studies of early cluster dissolution, show that a timescale of few Myr 
is long enough for a cluster to start evaporating and feed the surrounding field with low-mass stars 
due to violent relaxation induced by the expulsion of its residual star-forming gas \citep[e.g.,][]{kroupa08,
parmentier10}. This, considering the somewhat older age of the diffuse PMS population in the area between
the sub-clusters, may track the origin of these stars back at the central cluster 
NGC~602, which started evaporating almost immediately after formation. Indeed, the MST graph of Fig.~\ref{f:mst} 
for the whole detected sample of PMS stars, shows that these stars appear to be distributed 
almost symmetrically around NGC~602, as if they originated there. In addition, theoretical 
studies suggest that cool, clumpy young clusters dynamically mass segregate on a short timescale,  undergoing 
core collapse events, which can blow the clusters apart even with no need for gas expulsion
\citep{allison10}. Such processes may take place in the compact sub-clusters, still in their natal gas, which 
then would also contribute to the diffuse low-mass PMS population.  The same simulations showed that 
massive stars are often ejected from the cluster shortly after its formation with an average escape 
velocity of 2.5~km~s$^{-1}$. {It is interesting to note that this velocity is consistent with the upper 
limits for HI shell-expansion velocity from \cite{nigra08}.}
If we consider the upper age limit of 5~Myr for NGC~602, this escape velocity
would account for a distance of about 13~pc of a star from the cluster. Interestingly, this is almost identical 
to the projected radial distance, $\sim$~12~pc, of the most remote massive star, the O6 dwarf star \#8 in \cite{hutchings91} 
catalog. It is worth noting that 50\% of the hot massive stars in this catalog are located outside the boundaries of NGC~602,
as they are defined by the 2$\sigma$ isopleth in the density map of Fig.~\ref{f:scmap}.

Based on the discussion above we conclude that the central part of the region NGC~602/N90 is 
experiencing a multi-system active star formation for the last \lsim~5~Myr. The original event formed 
the cool, clumpy cluster NGC~602, which rapidly started dissolving into its immediate ambient, while becoming 
more compact at its center. Star formation continued taking place at the rim of the surrounding dusty nebular 
ring, in compact PMS sub-clusters, which are still 
half-embedded in their natal gas and bright in IR wavelengths. This process produced a diffuse 
inter-cluster PMS population originating from the clusters dissolution and introduced  an age-spread 
among the various PMS populations of the order of 2.5~Myr.

%The warmer and less substructured a cluster is initially, the less extreme its evolution.

\section{Concluding remarks}
\label{sec:discussion}

In this study we make use of the rich sample of PMS stars detected with ACS/WFC in the region
NGC~602/N90 in the SMC to investigate the clustering behavior of these stars, and derive 
conclusions on the origin of this behavior. We apply star-counts of PMS stars and we identify 14 
distinct  sub-clusters in terms of their high surface density {with respect to the average} background
stellar density in the region. The cluster NGC~602 is identified as the dominant stellar concentration 
in the region, located almost at the center of the ring-shaped nebula N90. {This cluster's stellar 
distribution is spatially asymmetric.} It is 
also found to be multi-peaked with few distinct stellar over-densities within its boundaries. We
verify that this multiplicity is real, since variable extinction is not high enough to produce it, and 
{photometric confusion can account for it only partly.} We find that the spatial  distribution of the PMS stars 
in the whole region is bimodal, with $\sim$~60\% of the total PMS population being included in the 
sub-clusters, and the remaining forming a diffuse population distributed in the inter-cluster area,  
covering the whole central part of the nebula with a higher concentration towards east.

We construct the CMDs of the detected clusters from all stars encompassed in the 2$\sigma$ isopleth
in the stellar density maps that defines the boundaries of each sub-cluster. {The age of young clusters 
cannot be determined directly from isochrone fitting of their low-mass PMS stars 
on the CMD. The reason is that observational and physical factors produce a broadening 
of these stars in the CMD that confuses the fitting process. As a consequence any isochrone fitting 
derives  unrealistic age spreads.} On the other hand the positions of intermediate-mass PMS stars 
are less affected by these factors, and therefore they are better tracers of the age of the
cluster. We were able to {establish upper limits on ages} from these stars only for the eight most populous sub-clusters, 
on the basis of isochrone fitting on the CMDs of the clusters, after the contamination by the 
background general field of the galaxy was statistically subtracted. We find that there is
a noticeable difference in the age of NGC~602 from that of its surrounding compact 
sub-clusters, of the order of $\sim$~2.5~Myr. The somewhat younger age of the compact 
sub-clusters is further supported by their half-embedded status, which makes them bright
in the IR, and some of them even coincide with candidate massive YSOs detected with
SST. On the other hand, most of the gas in NGC~602 seems to be expelled, {which   
attests to its more evolved nature.} An age estimate similar to that for NGC~602 was
found also for the diffuse PMS population from the corresponding CMD. 
{Neither the clustered nor the diffuse PMS population is found to be older than $\sim$~5~Myr.}

We apply the MST method and quantify the clustering behavior of PMS stars with the \Q\ parameter.
This application is not meaningful for clusters with less than 30 members, due to the derived
uncertainties, and therefore our analysis focuses only on the five most populous sub-clusters. We
find that almost all of them are centrally concentrated, with NGC~602 having $\Q = 0.8$, on the 
limit separating centrally concentrated from possible fractal clusters. These results do not change 
{when the UMS stars of the clusters are included in the \Q\ determination also.}  Judging from the smaller 
sizes and higher densities of the poorer sub-clusters, as well as from their tentative \Q\ values, we 
assess that they should be quite compact and not hierarchical. Considering the clustering behavior of 
the diffuse PMS population, we
evaluate it from all detected PMS stars with surface density corresponding to the average stellar 
density in the region, and therefore also including the detected sub-clusters. We find that $\Q\ = 0.85$
for the low-mass PMS stars and $\Q\ = 0.75$ when also the UMS are included. These numbers 
do not allow a clear characterization of the stellar concentration as centrally concentrated
or possibly fractal, but they show that the complete PMS stellar population of the central
area of NGC~602/N90, both clustered and diffuse, seem to behave as a coherent stellar 
concentration in terms of dynamical and stellar evolution.  

Taking into account the fact that young star clusters start to evaporate rapidly due to
the expulsion of their natal gas or due to early dynamical evolution, and based on the results discussed 
above  we propose a scenario that may explain the cluster formation in the region NGC~602/N90.
According to this scenario star formation took place in multiple high-density areas during
a period of 2.5~Myr, less than 5~Myr ago. The clumpy cluster NGC~602 at the center of the 
nebula {seems to have formed first}, and possibly it started dissolving, sending low-mass PMS 
stars in its surrounding ambient, shortly after its formation. More recent star formation occurred in individual compact
sub-clusters, which also may have fed the field with evaporating low-mass PMS stars. There are no 
indications if the two star formation events are actually sequential or not, but the small age spread
of the infant PMS stars and their clustering behavior indicates that the whole central part of
the region is dominated by a single loose stellar concentration, characterized by multiple 
high density peaks, which are bright in the IR and a central larger cluster which has most 
of its gas expelled. If the hot bright stars included in this system are considered then this 
stellar concentration may possibly classify as a stellar association in the making. {However, the validity 
of this claim depends on the stability of the system, which needs to be further investigated through its 
dynamical behavior.}

\acknowledgments
D.A.G. kindly acknowledges financial support by the German Aerospace 
Center (DLR) and the German Federal Ministry for Economics and 
Technology (BMWi)  through grant 50~OR~0908. S.S. was supported by 
DFG through grants SCHM~2490/1-1, KL~1358/5-2 
and SFB~881 ``The Milky Way System''. This work is based on observations 
made with the NASA/ESA {\em Hubble Space Telescope}, obtained from the 
data archive at the Space Telescope Science Institute (STScI). STScI is operated 
by the Association of Universities for Research in Astronomy, Inc.\ under NASA 
contract NAS 5-26555.

%{\it Facilities:} \facility{HST (ACS)}, \facility{Spitzer (IRAC)}, \facility{NTT ()}


\begin{thebibliography}{}

% \bibitem[Adams \& Myers(2001)]{adams01}
% Adams, F.~C., \& Myers, P.~C.\ 2001, \apj, 553, 744

%\bibitem[Allen et al.(2007)]{allen07}
%Allen, L., et al.\ 2007, in Protostars and Planets V, 
%ed.\ B. Reipurth, D. Jewitt, \& K. Keil (Tucson: Univ.\ Arizona Press), 361

\bibitem[Allison et al.(2010)]{allison10} 
Allison, R.~J., Goodwin, S.~P., Parker, R.~J., Portegies Zwart, S.~F., \& de Grijs, R.\ 2010, \mnras, 407, 1098 

% \bibitem[Ballesteros-Paredes et al.(2007)]{ballesteros07}
% Ballesteros-Paredes, J., Klessen, R.~S., Mac Low, M.-M., \& V\'{a}zquez-Semadeni, E. 2007,
% in Protostars and Planets V, ed.\ B. Reipurth, D. Jewitt, \& K. Keil (Tucson: Univ.\ Arizona Press), 63

%\bibitem[Bastian et al.(2007)]{bastian07} 
%Bastian, N., Ercolano, B., Gieles, M., et al.\ 2007, \mnras, 379, 1302 

%\bibitem[Bastian et al.(2009)]{bastian09} 
%Bastian, N., Gieles, M., Ercolano, B., \& Gutermuth, R.\ 2009, \mnras, 392, 868 

%\bibitem[Battinelli \& Demers(1998)]{battinelli98}
%Battinelli, P., \& Demers, S.\ 1998, \aj, 115, 1472

\bibitem[Battinelli \& Demers(1992)]{battinelli92}
Battinelli, P., \& Demers, S.\ 1992, \aj, 104, 1458 

% \bibitem[Bica \& Schmitt(1995)]{bica+schmitt95}
% Bica, E.~L.~D., \& Schmitt, H.~R. 1995, \apjs, 101, 41

% \bibitem[Bolatto et al.(2007)]{bolatto07} 
% Bolatto, A., et al. 2007, \apj, 655, 212

% \bibitem[Bonnell \& Bate(2006)]{BonnellBate2006}
% Bonnell, I.~A., \& Bate, M.~R.\ 2006, \mnras, 370, 488 
% 
% \bibitem[Bonnell et al.(2003)]{bonnell03}
% Bonnell, I.~A., Bate, M.~R., \& Vine, S.~G.\ 2003, \mnras, 343, 413 

%\bibitem[Billot et al.(2011)]{billot11} 
%Billot, N., Schisano, E., Pestalozzi, M., et al.\ 2011, \apj, 735, 28 

\bibitem[Bor\r{u}vka(1926)]{boruvka26}
Bor\r{u}vka, O. 1926, Pr{\'a}ce moravsk\'e p\v{r}\'{\i}rodov\v{e}deck\'e spole\v{c}nosti, 3, 37

\bibitem[Bourke et al.(2001)]{bourke01}
Bourke, T.~L., Myers, P.~C., Robinson, G., \& Hyland, A.~R.\ 2001, \apj, 554, 916 

% \bibitem[Brice\~{n}o et al.(2007)]{briceno07}
% Brice{\~n}o, C., Preibisch, T., Sherry, W.~H., Mamajek, E.~A., Mathieu, R.~D., Walter, 
% F.~M., \& Zinnecker, H. 2007,
% in Protostars and Planets V, ed.\ B. Reipurth, D. Jewitt, \& K. Keil (Tucson: Univ.\ Arizona Press), 345

\bibitem[Brott \& Hauschildt(2005)]{brott05} 
Brott, I., \& Hauschildt, P.~H.\ 2005, The Three-Dimensional Universe with Gaia, 576, 565 

\bibitem[Bruck(1976)]{bruck76} 
Bruck, M.~T.\ 1976, Occasional Reports of the Royal Observatory Edinburgh, 1, 1 

\bibitem[Carlson et al.(2007)]{carlson07}
Carlson, L.~R., et al.\ 2007, \apjl, 665, L109

\bibitem[Carlson et al.(2011)]{carlson11} 
Carlson, L.~R., et al.\  2011, \apj, 730, 78 

%\bibitem[Carpenter(2000)]{carpenter00}
%Carpenter, J.~M.\ 2000, \aj, 120, 3139

\bibitem[Cartwright \& Whitworth(2004)]{cw04}
Cartwright, A., \& Whitworth, A.~P. 2004, \mnras, 348, 589

%\bibitem[Casertano \& Hut(1985)]{casertano+hut85}
%Casertano, S., \& Hut, P. 1985, \apj, 298, 80

\bibitem[Castelli \& Kurucz(2003)]{castelli03} 
Castelli, F., \& Kurucz, R.~L.\ 2003, Modelling of Stellar Atmospheres, 210, 20P 

\bibitem[Chieffi \& Straniero(1989)]{chieffistraniero89} 
Chieffi, A., \& Straniero, O.\ 1989, \apjs, 71, 47 

\bibitem[Cignoni et al.(2009)]{cignoni09} 
Cignoni, M., et al.\  2009, \aj, 137, 3668 

\bibitem[Clark \& Bonnell(2005)]{clark05b}
Clark, P.~C., \& Bonnell, I.~A.\ 2005, \mnras, 361, 2 

\bibitem[D'Antona \& Montalb{\'a}n(2003)]{dantona03} 
D'Antona, F., \& Montalb{\'a}n, J.\ 2003, \aap, 412, 213 

\bibitem[Da Rio, et al.(2010)]{dario10} 
Da Rio, N., Gouliermis, D.~A., \& Gennaro, M.\ 2010, \apj, 723, 166

\bibitem[Degl'Innocenti et al.(2008)]{deglinnocenti08} 
Degl'Innocenti, S., Prada Moroni, P.~G., Marconi, M., \& Ruoppo, A.\ 2008, \apss, 316, 25 

% \bibitem[Clark et al.(2005)]{clark05a}
% Clark, P.~C., Bonnell, I.~A., Zinnecker, H., \& Bate, M.~R.\ 2005, \mnras, 359, 809
% 
% \bibitem[Clarke et al.(2000)]{clarke00}
% Clarke, C.~J., Bonnell, I.~A., \& Hillenbrand, L.~A.\ 2000, in Protostars and Planets IV, 
% ed.\ V. Mannings, A.~P. Boss, \& S.~S. Russell (Tucson: Univ.\ Arizona Press), 151

% \bibitem[Contursi et al.(2000)]{contursi00}
% Contursi, A., et al. 2000, \aap, 362, 310

% \bibitem[Crutcher(1999)]{crutcher99}
% Crutcher, R.~M.\ 1999, \apj, 520, 706 
% 
% \bibitem[Crutcher et al.(2008)]{crutcher08}
% Crutcher, R.~M., Hakobian, N., \& Troland, T.~H.\ 2008, ApJ, accepted (arXiv:0807.2862)

% \bibitem[Danforth et al.(2003)]{danforth03}
% Danforth, C.~W., Sankrit, R., Blair, W.~P., Howk, J.~C., \& Chu, Y.-H.
% 2003, \apj, 586, 1179

\bibitem[Davies et al.(1976)]{davies76} 
Davies, R.~D., Elliott, K.~H., \& Meaburn, J. 1976, \memras, 81, 89

% \bibitem[de Boer \& Savage(1980)]{deboer+savage80}
% de Boer, K.~S., \& Savage, B.~D. 1980, \apj, 238, 86

\bibitem[Dolphin(2000)]{dolphin00}
Dolphin, A. E. 2000, \pasp, 112, 1383


\bibitem[Elmegreen(2000)]{elmegreen00}
Elmegreen, B.~G.\ 2000, \apj, 530, 277 

\bibitem[Elmegreen(2010)]{elmegreen10} 
Elmegreen, B.~G.\ 2010, IAU  Symposium, 266, 3 

\bibitem[Efremov \& Elmegreen(1998)]{efelm98} 
Efremov, Y.~N., \& Elmegreen, B.~G.\ 1998, \mnras, 299, 588 

\bibitem[Elmegreen \& Scalo(2004)]{elmegrscalo04} 
Elmegreen, B.~G., \& Scalo, J.\ 2004, ARA\&A, 42, 211 

% \bibitem[Elmegreen(2006)]{elm06}
% Elmegreen, B.~G. 2006, in Globular Clusters, Guide to Galaxies, ed.\ T.
% Richtler et al.\ (Berlin: ESO/Springer), in press
% (astro-ph/0605519)

% \bibitem[Elmegreen(2007)]{elmegreen07}
% Elmegreen, B.~G.\ 2007, \apj, 668, 1064

% \bibitem[Elmegreen et al.(2000)]{elm00}
% Elmegreen, B.~G., Efremov, Y., Pudritz, R.~E., \& Zinnecker H. 2000,
% in Protostars and Planets IV, ed.\ V. Mannings, A.~P. Boss, \& S.~S. Russell 
% (Tucson: Univ.\ Arizona Press), 179

% \bibitem[Evans et al.(2006)]{evans06}
% Evans, C.~J., Lennon, D.~J., Smartt, S.~J., \& Trundle, C.\ 2006, \aap, 456, 623

\bibitem[Gennaro et al.(2011)]{gennaro11} 
Gennaro, M., Prada Moroni, P.~G., \& Tognelli, E.\ 2011, \mnras accepted (arXiv:1110.0852)

\bibitem[Girardi et al.(2002)]{girardi02} 
Girardi, L., Bertelli, G., Bressan, A., Chiosi, C., Groenewegen, M. A. T., Marigo, 
P., Salasnich, B., \& Weiss, A. 2002, A\&A, 391, 195

\bibitem[Goodman et al.(2009)]{goodman09} 
Goodman, A.~A., Rosolowsky, E.~W., Borkin, M.~A., Foster, J.~B., Halle, M., Kauffmann, J., 
\& Pineda, J.~E.\ 2009, Nature, 457, 63 

%\bibitem[Gouliermis(2005)]{hstprop10566G}
%Gouliermis, D.\ 2005, HST Proposal, 10566

%\bibitem[Gouliermis(2007)]{hstprop11547G}
%Gouliermis, D.\ 2005, HST Proposal, 11547

% \bibitem[Gouliermis et al.(2006)]{gouliermis06}
% Gouliermis, D.~A., Dolphin, A.~E., Brandner, W., \& Henning, T.\ 2006,
% \apjs, 166, 549

\bibitem[Gouliermis et al.(2007)]{gouliermis07}
%\bibitem[Gouliermis et al.(2007)]{gouliermis07}
Gouliermis, D.~A., Quanz, S.~P., \& Henning, T.\ 2007, \apj, 665, 306

\bibitem[Gouliermis et al.(2008)]{gouliermis08}
Gouliermis, D.~A., Chu, Y.-H., Henning, T., Brandner, W., Gruendl, R.~A., Hennekemper, E., \& Hormuth, F.\ 2008,
\apj, 688, 1050

\bibitem[Gouliermis et al.(2010)]{gouliermis10} Gouliermis, D.~A., 
et al.\ 2010, The Impact of HST on European Astronomy, 71 

\bibitem[Gouliermis et al.(2011)]{gouliermis11} 
Gouliermis, D.~A., Dolphin, A.~E., Robberto, M., et al.\ 2011, \apj, 738, 137 

% \bibitem[Hartmann et al.(2001)]{Hartmann01}
% Hartmann, L., Ballesteros-Paredes, J., \& Bergin, E.~A.\ 2001, \apj, 562, 852 

% \bibitem[Hatchell et al.(2005)]{hatchell05}
% Hatchell, J., Richer, J.~S., Fuller, G.~A., Qualtrough, C.~J., Ladd, E.~F., \& Chandler, C.~J.\ 2005, \aap, 440, 151

%\bibitem[Gutermuth et al.(2009)]{gutermuth09} 
%Gutermuth, R.~A., Megeath, S.~T., Myers, P.~C., Allen, L.~E., Pipher, J.~L., \& Fazio, G.~G.\ 2009, \apjs, 184, 18 

%\bibitem[Graham \&  Hell(1985)]{grahamhell85}
%Graham, R.L., \& Hell, P.\ 1985, Annals of the History of Computing, 7, 43

% \bibitem[Hennekemper et al.(2008)]{h08} 
% Hennekemper, E., Gouliermis, D.~A., Henning, T., Brandner, W., \& Dolphin, A.~E. 2008, \apj, 672, 914

% \bibitem[Heitsch et al.(2001)]{Heitschetal2001}
% Heitsch, F., Mac Low, M.-M., \& Klessen, R.~S.\ 2001, \apj, 547, 280 

\bibitem[Hartmann(2001)]{hartmann01} 
Hartmann, L.\ 2001, \aj, 121, 1030 

\bibitem[Henize(1956)]{henize56}
Henize, K. G. 1956, \apjs, 2, 315

\bibitem[Hodge(1985)]{hodge85}
Hodge, P.\ 1985, \pasp, 97, 530

% \bibitem[Hunter et al.(2008)]{hunter08}
% Hunter, I., et al.\ 2008, \aap, 479, 541

\bibitem[Hutchings et al.(1991)]{hutchings91}
Hutchings, J.~B., Cartledge, S., Pazder, J., \&
Thompson, I.~B.\ 1991, \aj, 101, 933

% \bibitem[Kennicutt(1988)]{kennicutt88}
% Kennicutt, R.~C., Jr.\ 1988, \apj, 334, 144

% \bibitem[Klessen(2001)]{Klessen2001b}
% Klessen, R.~S.\ 2001, \apjl, 550, L77 

\bibitem[Jeffries(2011)]{jeffries11} Jeffries, R.~D.\ 2011, 
Star Clusters in the Era of Large Surveys, in press (arXiv:1102.4752)

\bibitem[Kontizas et al.(1994)]{kontizas94} 
Kontizas, M., Kontizas, E., Dapergolas, A., Argyropoulos, S., \& Bellas-Velidis, Y.\ 1994, \aaps, 107, 77 

\bibitem[Klessen \& Burkert(2000)]{klessen00}
Klessen, R.~S., \& Burkert, A.\ 2000, \apjs, 128, 287

% \bibitem[Klessen et al.(2000)]{klessen00b}
% Klessen, R. S., Heitsch, F., \& Mac Low, M.-M. 2000, \apj, 535, 887

% \bibitem[Krumholz et al.(2007)]{krumholz07}
% Krumholz, M.~R., Klein, R.~I., \& McKee, C.~F.\ 2007, \apj, 656, 959

\bibitem[Kroupa(2008)]{kroupa08} 
Kroupa, P.\ 2008, The Cambridge N-Body Lectures, 760, 181 

\bibitem[Kruskal(1956)]{kruskal56}
Kruskal, J.~B. Jr. 1956, Proc.\ Amer.\ Math.\ Soc., 7, 48

\bibitem[Lada \& Lada(2003)]{lada03}
Lada, C.~J., \& Lada, E.~A.\ 2003, \araa, 41, 57

% \bibitem[Laney \& Stobie(1994)]{laney+stobie94}
% Laney, C.~D., \& Stobie, R.~S. 1994, \mnras, 266, 441 

% \bibitem[Le Coarer et al.(1993)]{lecoarer93}
% Le Coarer, E., Rosado, M., Georgelin, Y.~P., Viale, A., \& Goldes, G., 1993, \aap, 280, 365

\bibitem[Lee et al.(2005)]{lee05} Lee, H., Jackson, D.~C.,
Skillman, E.~D., Cannon, J.~M., Gehrz, R.~D., Polomski, E.,
\& Woodward, C.~E.\ 2005, Bulletin of the American Astronomical Society,
37, 1346

% \bibitem[Mac Low \& Ossenkopf(2000)]{maclow00b}
% Mac Low, M.-M., \& Ossenkopf, V.\ 2000, \aap, 353, 339 

\bibitem[Mac Low \& Klessen(2004)]{maclow+klessen04}
Mac Low, M.-M., \& Klessen, R.~S. 2004, Rev. Mod. Phys., 76, 125

% \bibitem[Massey et al.(1989)]{massey89}
% Massey, P., Parker, J.~W., \& Garmany, C.~D. 1989, \aj, 98, 1305

\bibitem[Massey et al.(2000)]{massey00}
Massey, P., Waterhouse, E., \& DeGioia-Eastwood, K.\ 2000, \aj, 119, 2214

\bibitem[McKee \& Ostriker(2007)]{mckee+ostriker07}
McKee, C. F., \& Ostriker, E.~C. 2007, \araa, 45, 565

\bibitem[McLaughlin \& Pudritz(1996)]{mclaughlin96}
McLaughlin, D. E., and Pudritz, R. E. 1996, ApJ, 457, 578

% \bibitem[Mouschovias(1991)]{Mou91}
% Mouschovias, T.~C.\ 1991, NATO ASIC Proc.~342: The Physics of Star Formation and Early Stellar 
% Evolution, 61 
% 
% \bibitem[Mouschovias \& Spitzer(1976)]{Mou76}
% Mouschovias, T.~C., \& Spitzer, L., Jr.\ 1976, \apj, 210, 326 

% \bibitem[Naz\'{e} et al.(2002)]{naze02}
% Naz\'{e}, Y., et al. 2002, \apj, 580, 225
% 
% \bibitem[Naz\'{e} et al.(2004)]{naze04} 
% Naz\'{e}, Y., Manfroid, J., Stevens, I. R., Corcoran, M. F., \& Flores, A.
% 2004, \apj, 608, 208

% \bibitem[Niemela et al.(1986)]{niemela86}
% Niemela, V. S., Marraco, H. G., \& Cabanne, M. L. 1986, \pasp, 98, 1133

\bibitem[Nigra et al.(2008)]{nigra08}
Nigra, L., Gallagher, J.~S., III, Smith, L.~J., Stanimirovi{\'c}, S., Nota, A., \& Sabbi, E.\ 2008, \pasp, 120, 972

\bibitem[Parmentier(2010)]{parmentier10} Parmentier, G.\ 2010, IAU 
Symposium, 266, 87

%\bibitem[Nota(2004)]{hstprop10248N}
%Nota, A.\ 2004, HST Proposal, 10248

%\bibitem[Nota(2005)]{hstprop10542N}
%Nota, A.\ 2005, HST Proposal, 10542

% \bibitem[Ossenkopf \& Mac Low(2002)]{Ossenkopf02}
% Ossenkopf, V., \& Mac Low, M.-M.\ 2002, \aap, 390, 307 
% 
% \bibitem[Ossenkopf et al.(2001)]{Ossenkopfetal2001}
% Ossenkopf, V., Klessen, R.~S., \& Heitsch, F.\ 2001, \aap, 379, 1005 

\bibitem[Prim(1957)]{prim57}
Prim, R.~C. 1957, Bell Syst.\ Tech.\ J., 36, 1389

% \bibitem[Reid et al.(2006)]{reid06}
% Reid, W.~A., et al.\ 2006, \mnras, 367, 1379

% \bibitem[Rochau et al.(2007)]{rochau07}
% Rochau, B., Gouliermis, D.~A., Brandner, W., Dolphin, A.~E., \& Henning,
% T. 2007, \apj, 664, 322

\bibitem[Rolleston et al.(1999)]{rolleston99}
Rolleston, W.~R.~J., Dufton, P.~L., McErlean, N.~D., \& Venn, K.~A.\ 1999,
\aap, 348, 728

% \bibitem[Rubio et al.(2000)]{rubio00}
% Rubio, M., et al. 2000, \aap, 359, 1139

% \bibitem[Sabbi et al.(2007)]{sabbi07}
% Sabbi, E., et al.\ 2007, \aj, 133, 44 

% \bibitem[Scalo(1985)]{scalo85}
% Scalo, J.~M. 1985, in Protostars and Planets II, ed.\ D.~C. Black
% \& M.~S. Matthews (Tucson: Univ.\ Arizona Press), 201

\bibitem[Sabbi et al.(2009)]{sabbi09} 
Sabbi, E., et al.\ 2009, \apj, 703, 721 

\bibitem[Schmalzl et al.(2008)]{schmalzl08} 
Schmalzl, M., Gouliermis, D.~A., Dolphin, A.~E., \& Henning, T.\ 2008, \apj, 681, 290 

\bibitem[Schmeja(2011)]{schmeja11} Schmeja, S.\ 2011, 
Astronomische Nachrichten, 332, 172 

% \bibitem[Schmeja \& Klessen(2004)]{sk04}
% Schmeja, S., \& Klessen, R.~S. 2004, \aap, 419, 405 

\bibitem[Schmeja \& Klessen(2006)]{sk06}
Schmeja, S., \& Klessen, R.~S. 2006, \aap, 449, 151

%\bibitem[Schmeja et al.(2008)]{skf08}
%Schmeja, S., Kumar, M.~S.~N., \& Ferreira, B. 2008, \mnras, 389, 1209

\bibitem[Schmeja et al.(2009)]{schmeja09}
Schmeja, S., Gouliermis, D.~A., \& Klessen, R.~S.\ 2009, \apj, 694, 367

\bibitem[Sherry, Walter, \& Wolk(2004)]{sherry04} 
Sherry, W.~H., Walter, F.~M., \& Wolk, S.~J.\ 2004, \aj, 128, 2316 

\bibitem[Shu et al.(1987)]{shu87}
Shu, F.~H., Adams, F.~C., \& Lizano, S.\ 1987, \araa, 25, 23 

% \bibitem[Simon et al.(2007)]{simon07}
% Simon, J. D., et al. 2007, \apj, 669, 327

\bibitem[Simon et al.(2000)]{simon00} 
Simon, M., Dutrey, A., \& Guilloteau, S.\ 2000, \apj, 545, 1034 

\bibitem[Stahler \& Palla(2005)]{sta+pal05}
Stahler, S. W., \& Palla, F. 2005, The Formation of Stars (Weinheim: Wiley-VCH)

\bibitem[Stanimirovic et al.(2000)]{stanimirovic00}
Stanimirovic, S., Staveley-Smith, L., van der Hulst,
J.~M., Bontekoe, T.~R., Kester, D.~J.~M., \& Jones, P.~A.\
2000, \mnras, 315, 791

\bibitem[Stassun et al.(2004)]{stassun04} 
Stassun, K.~G., Mathieu, R.~D., Vaz, L.~P.~R., Stroud, N., 
\& Vrba, F.~J.\ 2004, \apjs, 151, 357 

\bibitem[Steffen et al.(2001)]{steffen01} 
Steffen, A.~T., Mathieu, R.~D., Lattanzi, M.~G., et al.\ 2001, \aj, 122, 997 

\bibitem[Tognelli et al.(2011)]{tognelli11} 
Tognelli, E., Prada Moroni, P.~G., \& Degl'Innocenti, S.\ 2011, \aap, 533, A109 

\bibitem[Tognelli et al.(20011b)]{tognelli11b} 
Tognelli, E., Degl'Innocenti, S., \& Prada Moroni, P. G., 2011b, in preparation 
  
\bibitem[Vallenari et al.(2010)]{vallenari10} 
Vallenari, A., Chiosi, E., \& Sordo, R.\ 2010, A\&A, 511, A79 

\bibitem[Ventura et al.(1998)]{ventura98} 
Ventura, P., Zeppieri, A., Mazzitelli, I., \& D'Antona, F.\ 1998, \aap, 331, 1011 

\bibitem[Westerlund(1964)]{westerlund64} 
Westerlund, B.~E.\ 1964, \mnras, 127, 429 

% \bibitem[Staveley-Smith et al.(1997)]{staveleysmith97}
% Staveley-Smith, L., Sault, R.~J., Hatzidimitriou, D., Kesteven, M.~J., \& 
% McConnelle, D. 1997, \mnras, 289, 225

% \bibitem[V\'{a}zquez-Semadeni(2004)]{vazquez04}
% V\'{a}zquez-Semadeni, E. 2004, \apss, 292, 187

% \bibitem[V{\'a}zquez-Semadeni et al.(2003)]{Vazquez03}
% V{\'a}zquez-Semadeni, E., Ballesteros-Paredes, J., \& Klessen, R.~S.\ 2003, \apjl, 585, L131 

%\bibitem[von Hoerner(1963)]{vonhoerner63}
%von Hoerner, S. 1963, \zap, 57, 47

% \bibitem[Walborn et al.(2000)]{walborn00}
% Walborn, N.~R., et al. 2000, \pasp, 112, 1243

% \bibitem[Ye et al.(1991)]{ye91}
% Ye, T., Turtle, A.~J., \& Kennicutt, R.~C. Jr. 1991, \mnras, 249, 722

\end{thebibliography}
\end{document}